\documentclass[useAMS,usenatbib]{mn2e}

\def\apj{ApJ}                 % Astrophysical Journal
\def\aap{A\&A}                % Astronomy and Astrophysics
\def\mnras{MNRAS}             % Monthly Notices of the RAS
\def\nat{Nature}              % Nature
\def\PhyS{Physica Scripta}    %Physica Scripta
\def\RvMP{RevModPhys}         %Review of Modern Physics
\def\Aipc{AIP Conf. Proc.}     %AIP Conf. Proc.

\usepackage{epsf}
\usepackage{bm}
\title[Fields from a relativistic magnetic explosion]{Fields from a relativistic magnetic explosion}
\author[K.N. Gourgouliatos, D. Lynden-Bell]{K.N. Gourgouliatos\thanks {E-mail: kng22@ast.cam.ac.uk} \& D. Lynden-Bell\thanks{E-mail:
dlb@ast.cam.ac.uk}\\
Institute of Astronomy, University of Cambridge, Madingley Road CB3 0HA}
\begin{document}

\date{Accepted 2008 August 25. Received 2008 August 25; in original form 2008 July 16}

\pagerange{\pageref{firstpage}--\pageref{lastpage}} \pubyear{-}

\maketitle

%opening

\label{firstpage}
\begin{abstract}
Following Prendergast we study the relativistically expanding electromagnetic fields generated by an axisymmetric explosion of magnetic energy in a small volume. The magnetic field expands uniformly either within a cone or in all directions and it is therefore accompanied by an electric field. In the highly conducting plasma the charges move to annul the electric field in the frame of the moving plasma. The solutions presented are analytical and semi-analytical. We find that the time-scale for the winding up of the initial magnetic field is crucial, as short time-scales lead to strong radiant fields. Assuming a magnetic field of $10^{13}Gauss$ emerging from a magnetosphere of $10^{9}cm$ we end with a jet when confined by a pressure environment that falls more slowly than $r^{-4}$. The jet carries energy of $10^{51}erg$, which is mostly due to differential rotation at the base.  
 
\end{abstract}


\section{Introduction}

Observations of a wide variety of astronomical objects suggests the existence of magnetic fields in relativistic environments. There have been many studies of magnetic fields emanating from differentially rotating systems. Some of them confine themselves to the non-relativistic regime in which the displacement currents can be neglected. Even then few of them are analytic e.g. \cite{1976Natur.262..649L}, \cite{1994MNRAS...267..146L}, \cite{
	1994A&A...287..893S}, \cite{2006MNRAS.369.1167L}, while most of them are computational e.g. \cite{1970ApJ...161..541L}, \cite{1995MNRAS.277.1327B}, \cite{1997Natur.385..409O}. In the force-free case it was shown that the time evolution of the magnetic field arises solely from the time dependence of the boundary conditions so that the exact dynamical evolution can be calculated from the time dependent sequence of static models \citep{2006MNRAS.369.1167L}. Those in turn can be derived from the energy principle. However, that simplification depends on both the force-free condition and the neglect of the displacement currents which is only valid when the velocities are much less than $c$. Relativistic problems are harder as that approximation is invalid, so most studies are purely computational e.g. \cite{1992ApJ...394..459L}, \cite{2002MNRAS.336..759K}, \cite{2003ApJ...589..444G}, \cite{2005ApJ...620..878D} and \cite{2008MNRAS.388..551T} or semi-analytical e.g. \cite{1995ApJ...446...67C}.   

Despite this, \cite{2005MNRAS.359..725P} was able to find an exact solution to the relativistic MHD problem of a point magnetic explosion. He derived a time dependent relativistic analogue of the Grad-Shafranov equation that governs axially symmetric force-free MHD by assuming that the radial coordinate and the time since the explosion only appear in the dimensionless combination $v=r/ct$. Strictly speaking such equations are only valid when the length scale and $c \times$ the time scale of the region of the explosion are much smaller than $r$ and $ct$. The resulting Prendergast equation is non-linear but becomes linear in a special case. It was this special case that Prendergast studied in detail. However, he found that there are spherical nodes where the radial magnetic field is zero and these nodes occur before the highly relativistic regime $r \approx ct$ is reached. This has the unfortunate consequence that magnetic field lines emanating from small $r$ turn back before they reach the extremely relativistic region where the displacement currents are very important. The field lines in that region are an appendage unattached to their origin. Very similar effects are well known when one takes the analogous case, $\mathbf{j}= \alpha \mathbf{B}$ with $\alpha$ constant, in non-relativistic MHD in spherical coordinates. When $\alpha r$ becomes large, the system gives way to oscillating solutions with a series of nodes. This difficulty occurs because $\alpha^{-1}$ has dimension $L$ and the field has to vary on this fixed length scale even at large r. At large distances there is too much current for unit field and the smaller scale is then reflected in the scale of oscillation. 

It is known that non-linear ansatzes for the non-relativistic Grad-Shafranov equation can avoid this, which we now recognise as a bad consequence of a mathematically simple linear approximation which is not generally justified in the physics of the problem. We shall therefore study the non-linear Prendergast equation in all its glory!

We generalise the idea of force-free fields by considering a configuration where for each point there exists a Lorentz frame in which the electric field is zero and the current is along the magnetic field. The force density $\mathbf{f}$ in the frame fixed at the origin is then: 

\begin{eqnarray}
\mathbf{f}=\rho(\mathbf{E}+\mathbf{v} \times \mathbf{B})=0,
\end{eqnarray}

where $c\mathbf{v}(\mathbf{r},t)$ is the velocity of the moving frame. 
In this paper, we study such fields by solving the equation proposed by \cite{2005MNRAS.359..725P}. However, before stepping to a solution, we extract as much information as possible about the fields that is independent of the detailed form of the solution. Then we solve the equations semi-analytically as it impossible to achieve general analytical solutions. The solutions simplify in the non-relativistic limit and converge to those of \cite{1994MNRAS...267..146L}. There are some other cases that are  interesting and exactly soluble, namely the current-free magnetic dipole and the linear force-free field of Prendergast. The introduction of currents in the system allows the existence of a toroidal component of the field at the cost of making analytical solutions much harder, however it is still possible to design analytical solutions for this structure.

\section{Problem setup and solution strategy}

\subsection{Self-Similar form of Maxwell's Equations}

In accordance to Prendergast's formalism we consider an electromagnetic field configuration for which there is at least one frame of reference at each point of space where the electric field vanishes and the magnetic field is parallel to the electric current. We denote these frames and the physical quantities appearing there by a prime. We take each of these frames to move uniformly with velocity $v=r/ct$. Although each frame moves uniformly away from the origin and at each moment those frames further from the origin move faster, we are not using the expanding coordinates of cosmology. The frame of reference of the observer at the origin is the unprimed frame. The primed frames of reference move radially with respect to the unprimed frame under the scaled velocity $\mathbf{v}$

\begin{eqnarray}
\mathbf{v}= \frac{r}{ct}\mathbf{\hat{r}}.
\end{eqnarray}

Then we may write for the primed frame of reference:

\begin{eqnarray}
\mathbf{j}'=\alpha \mathbf{B}',
\end{eqnarray}

\begin{eqnarray}
\mathbf{E}'=0.
\end{eqnarray}

The magnetic field in the unprimed frame of reference is

\begin{eqnarray}
B_{r}=B_{r}', & B_{\theta}=\gamma B_{\theta}', & B_{\phi}=\gamma B_{\phi}',
\end{eqnarray}

where $\gamma=(1-v^{2})^{-1/2}$.
The electric field in the unprimed frame of reference is

\begin{eqnarray}
E_{r}=0, & E_{\theta}=v B_{\phi}, & E_{\phi}= -v B_{\theta}.
\end{eqnarray}

The electric current in the unprimed frame of reference is

\begin{eqnarray}
j_{r}=\frac{1}{\gamma}j_{r}'+v c \rho, & j_{\theta}=j_{\theta}', & j_{\phi}=j_{\phi}'.
\end{eqnarray}

By virtue of $\nabla \cdot \mathbf{B}= 0$, $\nabla \times \mathbf{E}=-\partial \mathbf{B}/\partial (ct)$, the self-similarity condition and (1) we re-express the fields in terms of the flux function $P$ and $T$. $P(r,\theta,t)$ is the magnetic flux through the cap of a sphere of radius $r$ subtending a semi-angle $\theta$ at the origin, and $T$ is related to the toroidal component $B_{\phi}$ and to the specific torque carried by field lines between $P$ and $P+dP$. Remembering that $v=r/ct$ and $\mu=\cos \theta$ we have

\begin{eqnarray}
\mathbf{B}=\frac{1}{2\pi r^{2}}\Big[- \frac{\partial P}{\partial \mu} \mathbf{\hat{r}}-\frac{v}{\sqrt{1-\mu^{2}}}\frac{\partial P}{\partial v} \bm{\hat{\theta}} +\frac{T}{\sqrt{1-\mu^{2}}}\bm{\hat{\phi}}\Big],
\end{eqnarray}

\begin{eqnarray}
\mathbf{E}=\frac{1}{2\pi r^{2}}\Big[\frac{v T}{\sqrt{1-\mu^{2}}} \bm{\hat{\theta}}+\frac{v^{2}}{\sqrt{1-\mu^{2}}}\frac{\partial P}{\partial v} \bm{\hat{\phi}}\Big].
\end{eqnarray}

Now we substitute the currents (7) and the electric and magnetic fields into $\nabla \times \mathbf{B}= \partial \mathbf{E}/\partial (ct) +4\pi/c \mathbf{j}$ and finally take into account charge conservation to get Prendergast's equation for the flux function 

\begin{eqnarray}
v^{2}(v^{2}-1)\frac{\partial^{2} P}{\partial v^{2}}+2v^{3}\frac{\partial P}{\partial v}- (1-\mu^{2}) \frac{\partial^{2} P}{\partial \mu^{2}}  \nonumber \\
=v\frac{d\beta}{dP}\Big(\frac{v \beta}{1-v^{2}}\Big).
\end{eqnarray}

In Prendergast's notation $\mathcal{H}(P)$ is $\frac{8 \pi^{2}}{c}\beta(P)$ and our $P$ and $T$ are $2\pi$ times his. 

The toroidal part $T$ is related to the source function $\beta$ by

\begin{eqnarray}
T=\frac{v}{1-v^{2}}\beta(P).
\end{eqnarray}

Our task is to determine the function $P$ by solving the partial differential equation (10). This is not a straightforward task; even under a numerical treatment of the problem, the solution depends on the form of the source function $\beta$ chosen. 
Our strategy for solving equation (10) is\\
1. to study the non-relativistic limit so as to see how it connects to earlier work.\\
2. to study the elementary case when there are no currents in the body of the plasma. The resulting self-similar solutions of Maxwell's equations become singular at r=ct if the fields at the origin are turned on suddenly. We demonstrate how these singularities disappear when the fields are turned on over a small finite time.\\
3. find what form of source function $\beta(P)$ allows solutions of (10) that are separable in $\mu$ and $v$. The special form of $\beta(P)$ required can not be made to vanish at $P=0$. However it gives an exact solution in regions where P is above a certain threshold, $P_{0}$, where $\beta(P_{0})=0=\beta'(P_{0})$. When $P$ is below that threshold the field joins smoothly to one with no currents.\\
4. The current-free solutions in this outer cocoon are not of product form but are found by analysis into suitable harmonics whose coefficients are computed to high accuracy.\\
5. Thus we end with a global solution that expands with the velocity of light within a narrow prescribed cone. It has high fields but no charges near $r=ct$. Below that we enter the current free cocoon, soon followed by the basic relativistic solution of product form which is accompanied by charges and currents.

\subsection{Properties of the field}

In this section we focus on the properties of the field. It may seem more natural to achieve a solution for equation (10) and then investigate the details of the field. However it is possible to obtain some physical understanding by studying general properties of the field that do not depend on the details of the solution.

We seek frames of reference where the electric field vanishes. Such frames are particularly interesting as they determine the velocity of the field and allow us to trace the time evolution of the magnetic field lines \citep{1994ppai.book.....S}. The setup of the problem gives an initial set of frames obeying this property, the uniformly expanding ones. Indeed the motion of the field lines may be viewed as a simple expansion. A charge $q$ that is placed at distance $r_{0}$ from the origin at time $t_{0}$ with velocity $\mathbf{v}_{0}=r_{0}/ct_{0}\mathbf{\hat{r}}$, does not feel any force, as there is no electric field and the magnetic field in this frame has no effect on a particle at rest in it. The particle will remain on the same magnetic field line as those are determined by $P$ which in turn is a function of $v=r/ct$. However, as the fields have components in the all three directions the velocity field of uniform expansion is not perpendicular to the magnetic field. Conventionally the velocity of field lines is chosen to be perpendicular to them. We examine other frames of reference where there is no electric field. In order to find such frames we use the following quadratic equation for the velocity field $c \mathbf{v}_{F}$ \citep{1975ctf..book.....L}, that describes frames of reference where the magnetic and the electric fields are parallel to each other 

\begin{eqnarray}
\frac{\mathbf{v_{F}}}{1+\mathbf{v_{F}}^2}= \frac{\mathbf{E} \times \mathbf{B}}{B^{2}+E^{2}}.
\end{eqnarray}

This velocity field is clearly perpendicular to $\mathbf{B}$. Given that the electric and the magnetic field by construction satisfy $\mathbf{E} \cdot\mathbf{B}=0$, which is a relativistic field invariant, we conclude that in a frame where $\mathbf{E'} \parallel \mathbf{B'}$ either the electric or the magnetic field vanishes. Thus equation (12) may be used to define the velocity of the magnetic field. However the general velocity $\mathbf{\widetilde{v}_{F}}$ of a frame in which $\mathbf{E'}$ and $\mathbf{B'}$ are parallel allows boosts in the  direction of the magnetic field and only determines the motion in the direction transverse to the magnetic field. The solution of the above quadratic equation gives the velocity we are interested in, however it is feasible to achieve a simpler formula for the motion of the field lines. Equation (1) yields that $\mathbf{E}=-\mathbf{v} \times \mathbf{B}$. Any velocity field of the form

\begin{eqnarray}
\mathbf{\widetilde{v}_{F}}=\frac{\mathbf{E} \times \mathbf{B}}{|\mathbf{B}|^{2}}+\lambda \mathbf{B},
\end{eqnarray}

for arbitrary $\lambda$, gives a set of frames of reference where the electric field vanishes. The Lorentz transformation of the electric field is given by \citep{1975clel.book.....J}

\begin{eqnarray}
\mathbf{E'}=\gamma (\mathbf{E}+\mathbf{v_{F}} \times \mathbf{B}) -\frac{\gamma^{2}}{\gamma+1}\mathbf{v_{F}}(\mathbf{v_{F}} \cdot \mathbf{E}).
\end{eqnarray}

Substituting (13) into (14) and working the cross products we obtain that $\mathbf{E'}=0$. Therefore equation (13) gives the general velocities that can be attributed to the field lines; we follow \cite{1975ctf..book.....L} in using the expression for $\lambda=0$ which gives a velocity perpendicular to the magnetic field. This is a solution to (12) provided $|\mathbf{E}|<|\mathbf{B}|$. By virtue of equations (8) and (9) we find that the usual velocity field is

\begin{eqnarray}
v_{F,r}=\frac{1}{4 \pi^{2} r^{4}|\mathbf{B}|^{2}}\Big[\frac{v^{3}}{1-\mu^{2}}\Big(\frac{\partial P}{\partial v}\Big)^{2}+v \frac{T^{2}}{1-\mu^{2}}\Big],
\end{eqnarray}

\begin{eqnarray}
v_{F,\theta}=-\frac{1}{4 \pi^{2} r^{4}|\mathbf{B}|^{2}}\frac{v^{2}}{\sqrt{1-\mu^{2}}}\frac{\partial P}{\partial \mu}\frac{\partial P}{\partial v},
\end{eqnarray}

\begin{eqnarray}
v_{F,\phi}=\frac{1}{4 \pi^{2} r^{4}|\mathbf{B}|^{2}}\frac{v T}{\sqrt{1-\mu^{2}}}\frac{\partial P}{\partial \mu}.
\end{eqnarray}

Relations (15)-(17) give the details of the motion of the field lines. At the first instance we can make some reasonable assumptions for the flux function $P$. A magnetic field consists of field lines rising from and returning to an imaginary spherical cap of radius $r_{0}$ and confined between angles $0$ and $\Theta$ corresponding to $\mu=1$ and $\mu=\mu_{0}$. In the case of a dipole field instead of a spherical cap we have the whole spherical surface, and $\mu_{0}=-1$. The flux function $P$ is zero at the boundaries and has a single maximum at $\mu=\mu_{1}$ where the field lines turn back. The toroidal component $T$ has to obey $T \to 0$ approaching the edges and normally the decrease has to be such so there is no singularity. The above assumption allows the determination of the signs of the derivatives appearing at relations (15)-(17). Therefore $\frac{\partial P}{\partial v} \leq 0$, as there cannot be any flux generated as the field moves upwards; $\frac{\partial P}{\partial \mu}>0$ for $\mu_{0}\leq \mu <\mu_{1}$; $\frac{\partial P}{\partial \mu}>0$ for $\mu_{1}<\mu\leq 1$ and $\frac{\partial P}{\partial \mu}=0$ at $\mu=\mu_{1}$.  

We conclude that the field lines viewed this way perform a complex motion that consists of a non-uniform radial expansion; a meridional motion where the parts of the field lines that lie in $\mu \in (\mu_{1},1)$ move in the direction of $-\bm{\hat{\theta}}$, those in $\mu \in (\mu_{0}, \mu_{1})$ in the direction of $\bm{\hat{\theta}}$; and an azimuthal motion where the parts of the field lines lying in $\mu \in (\mu_{0}, \mu_{1})$ move in the direction of $\bm{\hat{\phi}}$ and the rest in the direction of $-\bm{\hat{\phi}}$. The magnitude of the velocity depends on the details of the solution and cannot be determined by the previous arguments. 

\subsection{The non-relativistic limit}

The parameter $v$, that is the expansion velocity scaled to the speed of light gives an estimate of the importance of the relativistic terms for the problem. In the limit of $v<<1$ equation (10) becomes,
 
\begin{eqnarray}
v^{2}\frac{\partial^{2}P}{\partial v^{2}}+(1-\mu^{2})\frac{\partial^{2}P}{\partial \mu^{2}}=-v^{2}\beta\frac{d\beta}{dP}.
\end{eqnarray}

Equation (18) is similar to the Grad-Shafranov equation for axisymmetric magnetic fields \citep{1958, 1966}. There are many solutions of this equation in astrophysical context, \citep{1984smh..book.....P, 1994A&A...288.1012A, 1994MNRAS...267..146L}. Amongst these solutions the most appropriate to our case is that of \cite{1994MNRAS...267..146L} where the field is a self-similar quadruple. In our first paradigm we are going to apply this solution to a self-similar dipole. 

Assuming a self-similar solution we set $P=F_{max} v^{-l} f(\mu)$, where $F_{max}$ is a flux normalisation, formally the maximum flux at $v=1$, so that the angular part of the flux $f(\mu)\le 1$, therefore equation (18) becomes

\begin{eqnarray}
l(l+1)v^{-l}F_{max}f(\mu)+(1-\mu^{2})F_{max}v^{-l}f(\mu)''=-v^{2}\beta\frac{d\beta}{dP}.
\end{eqnarray} 

In order to achieve self-similar solutions we need $\beta d\beta /dP \propto v^{-l-2}$. With $P$ a product of a function of $v$ and a function of $\mu$, $\beta(P)$ can only be proportional to a power of $v$ if $\beta$ itself is a power of $P$, since $P$ is proportional to $v^{-l}$ the required power is given by

\begin{eqnarray}
\beta =c_{1}P^{1+1/l},
\end{eqnarray}

a more detailed derivation is given in \cite{1994MNRAS...267..146L}. Setting $c_{0}=c_{1}F_{max}^{1/l}$, equation (19) reduces to

\begin{eqnarray}
l(l+1)f(\mu)+(1-\mu^{2})f(\mu)''=-c_{0}^{2}(1+\frac{1}{l})f^{2/l+1}.
\end{eqnarray}

To solve equation (21) in a sphere we apply the boundary conditions $f(1)=f(-1)=0$, and determine $c_{0}$ by the condition that $f$ has a single maximum $f(\mu_{max})=1$. These conditions are sufficient to solve the equation for any $l$. For $l=1$ the solution is $f(\mu)=1-\mu^{2}$ and $c_{0}=0$, for $l$ small the solution of (21) takes the form $f=1-l/(l+1)\ln\cosh((1+l^{-1})c_{0}\mu)$ and in general for intermediate values it is solved numerically. Therefore the fields take the following form in this non-relativistic limit

\begin{eqnarray}
\mathbf{B}=\frac{F_{max}}{2\pi r^{2}v^{l}}\Big[-f'\mathbf{\hat{r}}+\frac{lf}{\sqrt{1-\mu^{2}}}\bm{\hat{\theta}}+\frac{c_{0}f^{1+1/l}}{\sqrt{1-\mu^{2}}}\bm{\hat{\phi}}\Big],
\end{eqnarray}

\begin{eqnarray}
\mathbf{E}=\frac{F_{max} v}{2\pi r^{2}v^{l}\sqrt{1-\mu^{2}}}\Big[c_{0}f^{1+1/l}\bm{\hat{\theta}}-lf\bm{\hat{\phi}}\Big].
\end{eqnarray}

The motion of the field lines can be found by equations (15)-(17) by applying the fields given at (22) and (23).

\begin{eqnarray}
\mathbf{v_{F}}=\frac{v}{(1-\mu^{2})f'^{2}+l^{2}f^{2}+c_{0}^{2}f^{2/l+2}}[(l^{2}f^{2}+c_{0}^{2}f^{2+2/l})\mathbf{\hat{r}} \nonumber \\
+(1-\mu^2)^{1/2}lff'\bm{\hat{\theta}}+(1-\mu^2)^{1/2}f'f^{1+1/l}\bm{\hat{\phi}}].
\end{eqnarray}
 
This case is different to the Grad-Shafranov equation, as in the latter the field only depends on $r$ whereas here the field depends on $r/ct$. However, a snapshot of the configuration for any given $t$ satisfies the Grad-Shafranov equation. 

\section{Relativistic Solutions}

\subsection{Current-Free Solutions}

When there are no source terms on the right, equation (10) gives simple relativistic solutions with no toroidal magnetic field component.

In this section we study this case; we show that one solution reduces to a linearly increasing magnetic dipole, where the field is zero outside a sphere expanding at the speed of light.

With $\beta=0$ equation (10) reduces to the simple form

\begin{eqnarray}
v^{2}(v^{2}-1)\frac{\partial^{2} P}{\partial v^{2}}+2v^{3}\frac{\partial P}{\partial v}- (1-\mu^{2}) \frac{\partial^{2} P}{\partial \mu^{2}}=0.
\end{eqnarray}

By using Ogilvie's transformation $u=1/v$ equation (25) becomes

\begin{eqnarray}
(1-u^2)\frac{\partial ^{2} P}{\partial u^{2}}-2u\frac{\partial P}{\partial u} -(1-\mu^{2})\frac{\partial^{2}P}{\partial \mu^{2}}=0.
\end{eqnarray}

We use the technique of separation of variables; let $P(u,\mu)=R(u)M(\mu)$ and dash  denotes differentiation with respect to $u$, equation (26) becomes

\begin{eqnarray}
\frac{(1-u^{2})R''-2uR'}{R}=\frac{(1-\mu^{2})\frac{\partial^{2}M}{\partial \mu^{2}}}{M}=-l(l+1).
\end{eqnarray}  

Equation (27) for $R(u)$ is the Legendre differential equation, therefore the solution is a linear combination of $R(u)=c_{a}P_{l}(u)+c_{b}Q_{l}(u)$, that has to be finite in the interval $u \in [1,\infty)$, therefore $c_{b}=0$, as $Q_{l}(u)$ becomes infinite at $u=1$.

The angular part can be solved analytically using the following transformation \citep{1992ApJ...391..353W}, $M=(1-\mu^2)\frac{dM_{1}}{d\mu}$, therefore the second equality of (27) becomes:

\begin{eqnarray}
(1-\mu^{2})M_{1}''-2\mu M_{1}'+l(l+1)2M_{1}=0.
\end{eqnarray}  
 
This is again the Legendre differential equation, and the solution for the angular part of $P$ is $M(\mu)=(1-\mu^{2})P_{l}'(\mu)$. Therefore the flux function of the field is

\begin{eqnarray}
P(u, \mu)= \Big\{ \begin{array}{ll} 
F_{max}(1-\mu^{2})P_{l}'(\mu)P_{l}(u) & u \geq 1\\
0                                   & u < 1
\end{array} 
\end{eqnarray}

\subsubsection{The vacuum expanding dipole}

Let us consider $l=1$, that corresponds to a field where a dipole of magnetic moment $\mathbf{M}=M_{0}t\mathbf{\hat{z}}$ is initiated at $t=0$. Following the formalism of \cite{1952} for the Hertzian dipole modified for a magnetic dipole we can express the electric and the magnetic fields using the quantity $\mathbf{\Pi}$ which is a vector pointing at the axis of dipole $\mathbf{\hat{z}}$ and is a function of position and time; dot denotes derivation with respect to $ct$ 

\begin{eqnarray}
\mathbf{E}=-\nabla \times \dot{\mathbf{\Pi}},
\end{eqnarray}

\begin{eqnarray}
\mathbf{B}=-\ddot{\mathbf{\Pi}}+\nabla(\nabla \cdot \mathbf{\Pi})
\end{eqnarray}

It is shown in appendix A that for any function $D(ct-r)$ 

\begin{eqnarray}
\mathbf{\Pi}=\frac{D(ct-r)}{r}\mathbf{\hat{z}},
\end{eqnarray} 

and the fields satisfy Maxwell's equations. The fields are:

\begin{eqnarray}
\mathbf{E}=-\frac{\sin \theta}{r}\Big(\ddot{D}+\frac{\dot{D}}{r}\Big)\bm{\hat{\phi}}
\end{eqnarray}

\begin{eqnarray}
\mathbf{B}=\frac{2\cos \theta}{r^{2}}\Big(\dot{D}+\frac{D}{r}\Big)\mathbf{\hat{r}}+\frac{\sin \theta}{r}\Big(\ddot{D}+\frac{\dot{D}}{r}+\frac{D}{r^{2}}\Big)\bm{\hat{\theta}}.
\end{eqnarray}

Therefore, we can evaluate the electric and the magnetic fields for a linearly increasing dipole that is switched on at $t=0$ and lies at $r=0$, it corresponds to $D=m(ct-r)H(ct-r)$, where $H$ is the Heaviside function which is unity for argument greater than zero and zero otherwise. We divide the space into three regions. The first one (I) is $r<ct$; the second (II) is the surface $r=ct$ and the third (III) is $r>ct$. In the first region there is an electromagnetic field because of the dipole; in the the third region the field is zero, as the message of switching on the dipole has not arrived yet; the second region is the horizon surface of discontinuity. Therefore in region (I) the fields are:
 
\begin{eqnarray}
\mathbf{E}=-m\frac{\sin\theta}{r^{2}}\bm{\hat{\phi}},
\end{eqnarray} 

\begin{eqnarray}
\mathbf{B}=mc\frac{t}{r^{3}}(2\cos\theta \mathbf{\hat{r}}+\sin \theta \bm{\hat{\theta}}).
\end{eqnarray}

Since the fields are related to derivatives of $D$ and the first derivative of $D$ is discontinuous at $r=t$ we expect singularities on the surface $r=ct$ (II). Indeed there are singular surface fields. These infinities we should have expected because in the analogous electric dipole case velocities are suddenly imposed on the charges to make the linearly growing dipole. Thus initial accelerations are infinite so the power radiated should be singular.  

\begin{eqnarray}
\mathbf{E}=-m\frac{\sin \theta}{r}\Big(\delta(ct-r)+\frac{H(ct-r)}{r}\Big)\bm{\hat{\phi}},
\end{eqnarray}

\begin{eqnarray}
\mathbf{B}=\frac{mc}{r}\Big(\frac{2t \cos \theta H(ct-r)}{r}\mathbf{\hat{r}}+ \nonumber \\
\sin \theta(\delta(ct-r)+\frac{tH(ct-r)}{r^{2}})\bm{\hat{\theta}}\Big).
\end{eqnarray}

These fields are unacceptable. In region (III) there are no fields at all. To avoid the singularities in (II) we take into account the finite time needed to accelerate the dipole to a given growth rate. To do this we smooth by averaging the fields over a small spread $\tau_{0}$ of switch-on times. Thus region (I) now extends $r<c\big(t+\frac{\tau_{0}}{2}\big)$, region (II) spreads into a spherical shell $c(t-\tau_{0})<r<ct$ and (III) $r>c\big(t+\frac{\tau_{0}}{2}\big)$. $\bar{D}(ct-r)$ becomes: 

\begin{eqnarray}
\bar{D}= \Big\{ \begin{array}{lll} 
m(ct-r)   & \frac{r}{c}+\frac{\tau_{0}}{2}<t & (I) \\
\frac{m}{c\tau_{0}}(c(t+\frac{\tau_{0}}{2})-r)^{2} & \frac{r}{c}-\frac{\tau_{0}}{2}<t<\frac{r}{c}+\frac{\tau_{0}}{2} & (II)\\
0                        & t < \frac{r}{c}-\frac{\tau_{0}}{2} & (III)
\end{array} 
\end{eqnarray}
 
This substitution increases the complexity of the formulae for the fields but removes the  singularity at the horizon and allows us to study in greater detail the structure of the field near the horizon and how it depends on the time-scale for the dipole to establish a linear behaviour. The fields corresponding to $\bar{D}$ are given by putting $\bar{D}$ into (33) and (34) 

\begin{eqnarray}
\bar{E_{\phi}}= \Bigg\{ \begin{array}{ll} 
-m\frac{\sin \theta}{r^{2}}   & (I) \\
-m\frac{\sin \theta}{c\tau_{0}r}\Big(2+\frac{2(c(t+\frac{\tau_{0}}{2})-r)}{r}\Big)& (II)\\
0                        & (III)
\end{array} 
\end{eqnarray}

\begin{eqnarray}
\bar{B_{r}}= \Bigg\{ \begin{array}{ll} 
2mct\frac{\cos\theta}{r^{3}}  & (I) \\
\frac{2m\cos \theta}{c\tau_{0}r^{2}}\Big(\frac{(c(t+\frac{\tau_{0}}{2}))^{2}}{r}-r\Big) & (II)\\
0                       & (III)
\end{array} 
\end{eqnarray}

\begin{eqnarray}
\bar{B_{\theta}}= \Bigg\{ \begin{array}{ll} 
\frac{mct\sin \theta}{r^{3}}  & (I) \\
\frac{m\sin\theta}{c\tau_{0}r}\Big(\big(\frac{c(t+\frac{\tau_{0}}{2})}{r}\big)^{2}+1\Big)& (II)\\
0                       & (III).
\end{array} 
\end{eqnarray}

These are perfectly regular and acceptable.

The higher order self-similar multipoles are given in Appendix B.

\subsection{Solutions containing currents}

In the previous sections we have given analytical expressions consisting of magnetic fields that lie on the meridional planes and electric fields that are azimuthal; such fields do not contain currents, or charge densities and each arises from a growing multipole at the origin. In this section we seek analytical solutions for electromagnetic fields in which the magnetic field has an azimuthal component, the electric has a meridional component and there are charge and current densities in the space. 

We need to solve (10), Prendergast's equation. We seek a solution that is a product of a function depending on $v$ and one depending on $\mu$; using the usual transformation $v=1/u$ equation (10) reduces to

\begin{eqnarray}
\frac{\partial}{\partial u}\Big[(u^2-1)\frac{\partial P}{\partial u}\Big]+(1-\mu^{2})\frac{\partial^{2}P}{\partial \mu^{2}}=-\frac{1}{u^{2}-1}\beta \frac{d \beta}{d P}.
\end{eqnarray}

We multiply by $(u^{2}-1)/P$ and look for solutions in the form of $P=g(u)F(\mu)$

\begin{eqnarray}
\frac{u^{2}-1}{g}\frac{d}{du}\Big[(u^{2}-1)\frac{dg}{du}\Big]+(u^{2}-1)\frac{1-\mu^{2}}{F}\frac{d^{2}F}{d\mu^{2}}= \nonumber \\
=-\frac{\beta}{P}\frac{d\beta}{dP}.
\end{eqnarray}

We set:

\begin{eqnarray}
U(u)=\frac{u^{2}-1}{g}\frac{d}{du}\Big[(u^{2}-1)\frac{dg}{du}\Big],\\
M(\mu)=\frac{1-\mu^{2}}{F}\frac{d^{2}F}{d\mu^{2}},\\
\frac{\beta}{P}\frac{d\beta}{dP}=Q(P),
\end{eqnarray}

so the equation takes the form:

\begin{eqnarray}
U(u)+(u^{2}-1)M(\mu)=-Q(P),
\end{eqnarray}

operating with $F/F'\partial / \partial \mu$ we find

\begin{eqnarray}
(u^{2}-1)\frac{FM'}{F'}=-PQ',
\end{eqnarray}

which is of the form $A(g)B(F)=C(gF)$, with $A(g(u))=u^2-1$ and $C(P)=-PQ'$. Taking logs and operating with $gd/dg$ we find that (i) $d\ln A/d\ln g=d\ln C/d\ln P$ or alternatively (ii) $C=B=0$ which leads to Prendergast's linear model $\beta(P)\propto P$. Henceforward we consider only case (i), $F$ is not constant since $P$ can not be a function of $g$ alone, so both $d\ln A/d\ln g$ and $d\ln C/d\ln P$ must equal the same constant $2/l$. This is a generalisation of the $l$ used before in the current free case. Thus without loss of generality $g=A^{l/2}=(u^{2}-1)^{l/2}$ and $C=-PQ'=-KP^{2/l}$ where $K$ is constant. Finally from (77) $FM'/F'=-KP^{2/l}/(u^{2}-1)=-KF^{2/l}$, hence: $M'=-KF^{2/l-1}F'$, $M=-(\frac{1}{2}Kl F^{2/l}+c)$, $Q'=KP^{2/l-1}$ so $Q=\frac{1}{2}Kl P^{2/l}+c'$, our basic equation (44) now reads:

\begin{eqnarray}
(u^{2}-1)^{1-l/2}\frac{d}{du}\Big[(u^{2}-1)\frac{d}{du}(u^{2}-1)^{l/2}\Big]{\nonumber}\\
-(u^{2}-1)(\frac{1}{2}Kl F^{2/l}+c)=-(\frac{1}{2} Kl P^{2/l}+c'),
\end{eqnarray}

but

\begin{eqnarray}
(u^{2}-1)^{1-l/2}\frac{d}{du}\Big[lu (u^{2}-1)^{l/2}\Big]=l(u^{2}-1)+l^{2}u^{2},
\end{eqnarray}

so equation (50) reduces to a linear equation in $u^{2}$. Comparing coefficients of $u^{2}$

\begin{eqnarray}
l(l+1)-(\frac{1}{2}Kl F^{2/l}+c)=-\frac{1}{2}Kl F^{2/l}
\end{eqnarray}

and from the coefficients of $u^{0}$ in (50)

\begin{eqnarray}
-l +\frac{1}{2}K l F^{2/l}+c=\frac{1}{2}K l F^{2/l}-c'.
\end{eqnarray}

Evidently we must choose $c=l (l+1)$, so the first is satisfied and $c'=-l^{2}$ so the second is too. However in the non-relativistic approximation $u^{2}>>1$ so then the second equation is all negligible compared with the first. 

From the definition of $M$ we have:

\begin{eqnarray}
\frac{1-\mu^{2}}{F}\frac{d^{2}F}{d \mu^{2}}=-\frac{1}{2}KlF^{2/l}-l(l+1),
\end{eqnarray}

so 

\begin{eqnarray}
(1-\mu^{2})F''+l(l+1)F=\frac{1}{2}KlF^{1+2/l},
\end{eqnarray}

and by normalising $f=F/F_{max}$, so $f \leq 1$

\begin{eqnarray}
(1-\mu^{2})f''+l(l+1)f=-\frac{1}{2}KlF_{max}^{2/l}f^{1+2/l},
\end{eqnarray}

which is the same equation as that studied by \cite{1994MNRAS...267..146L}, who found it in non-relativistic MHD. We choose the solutions obeying boundary conditions that $f$ is zero at $\theta=0$, ($\mu=1$) and at any chosen outer boundary $\Theta$, ($\mu_{0}$). The constant $C=\frac{1}{2}KlF_{max}^{2/l}$ is then determined in such a way that $f$ is one at its maximum. 
For small $l$ and $\mu=1-\frac{1}{2}\theta^{2}$ the solution for $f$ is

\begin{eqnarray}
f=1-\frac{1}{\nu}\ln\cosh\big[\nu g\big(\frac{2\theta^{2}}{\Theta^{2}}-1\big)\big],
\end{eqnarray}

where $\nu=1+1/l$ and $g$ is close to 1 and given by

\begin{eqnarray}
g=\frac{1}{\nu}\cosh^{-1}e^{\nu}=1+\frac{1}{\nu}\ln\big[1-\big(\frac{1-\sqrt{1-e^{-2\nu}}}{2}\big)\big] \nonumber\\
\simeq 1+\frac{1}{\nu}\ln2.
\end{eqnarray}

The value of $C$ associated with this solution is

\begin{eqnarray}
\frac{1}{2}KlF^{2/l}_{max}=C=\frac{8g^{2}}{\nu \Theta^{2}}.
\end{eqnarray}

Equation (57) gives solutions which are symmetric in $\theta^{2}$ around their maximum $\Theta^{2}/2$. This is the case for $l=0$, when $l$ is small but not zero the maximum is slightly displaced. A more accurate formula taking into account this displacement is

\begin{eqnarray}
f=1- \frac{1}{\bar{\nu}} \ln \cosh \Big(\frac{\bar{C}\bar{\nu}}{\sin \theta_{1}}\{\mu-\mu_{1} \nonumber \\
+\frac{\Theta^{2}}{12 \bar{\nu}^{2}} \ln \cosh[\frac{\bar{C}\bar{\nu}}{\sin \theta_{1}}(\mu-\mu_{1})]\}\Big),
\end{eqnarray}

where $\bar{\nu}=\nu-\frac{\Theta^{2}}{16}\big(1-\frac{2\ln 2}{\nu}\big)$, $\bar{C}=\frac{2 \sqrt{2}}{\Theta}\big(1+\frac{\ln2}{\nu}\big)$, $\theta_{1}=\frac{\Theta}{\sqrt{2}}-\frac{1}{6(\nu +2\ln2)}$, $\mu_{1}=1-\frac{1}{2}\theta_{1}^{2}$ and $C=\bar{C}^{2} \bar{\nu}-l^{2}\nu$, for details on the derivation see \cite{2006MNRAS.369.1167L} (appendix).

The remaining equation comes from the definition of $Q$ (47), 

\begin{eqnarray}
\beta^{2}=\frac{Kl^{2}}{2(l+1)}P^{2(1+1/l)}-l^{2}P^{2}+c_{2},
\end{eqnarray}

where $c_{2}$ is constant. If $l>0$ then $\beta^{2}$ has a minimum at $P_{0}=(2l/K)^{l/2}$. We choose $c_{2}$ so that this minimum is zero and take $\beta$ to be zero for $P<P_{0}$ so 

\begin{eqnarray}
\beta^{2}= \Bigg\{ \begin{array}{l} 
l^{2}P_{0}^{2}\Big\{\frac{l}{l+1}\big[(P/P_{0})^{2+2/l}-1\big]-\big[(P/P_{0})^{2}-1\big]\Big\}\\
0, P<P_{0}
\end{array} 
\end{eqnarray}

Had we chosen a lower value of $c_{2}$ then $\beta^{2}$ would have been negative for values of $P$ near $P_{0}$. However $\beta^{2}$ must be positive. An unfortunate consequence of choosing the upper expression (62) for $\beta^{2}$ for all $P$ is that $\beta^{2}\neq 0$ when $P=0$. This entails an infinite $B_{\phi}$ near the axis which corresponds to a line current there. To avoid this in a continuous and differentiable way we choose the upper form (62) only where $P\geq P_{0}$. Elsewhere we choose $\beta\equiv 0$, the current-free solution. We avoid the use of the linear form $\beta \propto P$ because those solutions have an infinity of oscillations in the sign of $P$ as they approach $u=1$ (see Prendergast 2005). So they contain fields that are disconnected from the origin. 

Thus, our solution of the relativistic Prendergast equation is

\begin{eqnarray}
P=F_{max}f(\mu)(u^{2}-1)^{l/2}; & P\geq P_{0},
\end{eqnarray}

where $f$ is given by (57) or more accurately by (60).

In what follows we determine the fringing field in the cocoon region $P<P_{0}$. We choose the boundary to be a cone with a total opening angle (not semi-angle) of less than $22^{o}$. This is the case of greatest importance for applications to $\gamma$-ray bursts, however we should point out that exactly similar methods work for any cone or even for the whole sphere. It is just that Legendre polynomials or more generally Legendre functions of $\mu$ replace the simpler Bessel functions of the narrow cone cases.

\subsubsection{Solution in narrow cones}

Since our main applications will be the jets and $\gamma$-ray bursts it makes sense to study solutions within conical boundaries with quite narrow cones $\theta \leq \Theta$. Provided the total opening angle (not the semi-angle) of the cone is less than $22^{o}$ the actual opening angle does not affect the form of the solution whereas at larger angles it does. Thus there is a real advantage in studying the small angle case which allows us to replace high order Legendre functions of angles by Bessel functions of order 1. The $11^{o}$ limit comes from our replacement of $(1+\mu)=(1+\cos\theta)\approx 2-\frac{1}{2}\theta^{2}$ by 2. The fractional error in neglecting $\frac{1}{2}\theta^{2}$ as compared with 2 is $\frac{1}{4}\theta^{2}$ which is less than $1\%$ for $\theta<1/5 \simeq 11^{o}$

In the region where $P<P_{0}$, $\beta^{2}$ is zero so the solutions of (43) that we need, have no currents, however they are solutions with a very awkward boundary on $P=P_{0}$. In place of trying to fit a current-free solution on that boundary to our known solution within the region with $P<P_{0}$ we adopt the current-free solutions within the whole cone as our complete set. We expand the known right hand side of equation (43) in terms of our Bessel functions of angle within the cone. We then get a known function of $u$ as a source term on the right hand side of each Bessel component of the equation (43). We solve for the radial functions of $u$ which are the coefficients of the different Bessel components, each with its own known source term. 

Under the assumption of narrow opening angle and in the absence of currents the angular part of (43) becomes 
 
\begin{eqnarray}
\theta\frac{d}{d\theta}\Big(\frac{1}{\theta}\frac{d\tilde{J}}{d\theta}\Big)+\nu(\nu+1)\tilde{J}=0,
\end{eqnarray}   

setting $\tilde{J}=\theta\frac{dJ}{d \theta}$ the above equation becomes:

\begin{eqnarray}
\theta \frac{d}{d\theta}\Big[\frac{1}{\theta}\frac{d}{d \theta}\Big(\theta \frac{dJ}{d \theta}\Big)+\nu(\nu+1)J\Big]=0,
\end{eqnarray}

so

\begin{eqnarray}
\frac{1}{\theta}\frac{d}{d \theta}\Big(\theta \frac{dJ}{d \theta}\Big)+\nu(\nu+1)J=const.
\end{eqnarray}

So far, we have only defined J up to an arbitrary additive constant which is eliminated in $\tilde{J}$ so we absorb the constant into $J$ and get:

\begin{eqnarray}
\frac{1}{\theta}\frac{d}{d \theta}\Big(\theta \frac{dJ}{d \theta}\Big)+\nu(\nu+1)J=0.
\end{eqnarray}

The solution is $a J_{0}(k\theta)+b Y_{0}(k\theta)$, where $\nu(\nu+1)=k^{2}$. If we want a solution regular at $\theta=0$ we must omit the $b$ term so $J=aJ_{0}(k\theta)$, and 

\begin{eqnarray}
\tilde{J}=ak\theta\frac{dJ_{0}(k\theta)}{d(k\theta)}=-azJ_{1}(z),
\end{eqnarray}

where $z=k\theta$. 

In order that $\tilde{J}$ should vanish at $\Theta$ we need $J_{1}(k\Theta)=0$. The solutions for $k$ are $k_{s}=j_{s}/\Theta$ where $j_{1}=3.83$, $j_{2}=7.01$, $j_{3}=10.17$ etc p 409 of A \& S, and $j_{s}$ is the s-th zero of $J_{1}(z)$ for $z>0$, called $j_{1s}$ in A \& S.

Now any function $F(\theta)$, $F(0)=F(\Theta)=0$ can be expanded as a series. 

\begin{eqnarray}
F(\theta)=\sum_{s}F_{s}J_{1}(j_{s}\theta/\Theta).
\end{eqnarray}

The orthogonality relation is A \& S 11.4.5 

\begin{eqnarray}
\int_{0}^{1}tJ_{1}(j_{s}t)J_{1}(j_{n}t)dt=\frac{1}{2}[J_{1}'(j_{n})]^{2}\delta_{sn},
\end{eqnarray}

To find $F_{s}$ we multiply by $\theta/\Theta J_{1}(j_{n}\theta /\Theta)$ and integrate:

\begin{eqnarray}
\int_{0}^{1}[\frac{\theta}{\Theta}J_{1}(j_{n}\theta /\Theta) \sum_{s}F_{s}J_{1}(j_{s}\theta /\Theta)]d\Big(\frac{\theta}{\Theta}\Big)= \nonumber \\
=\frac{1}{2}[J_{1}'(j_{n})]^{2}F_{n}.
\end{eqnarray}

So the $F_{n}$ corresponding to any $F$ can by found by integration and a knowledge of the $j_{n}$ and $J_{1}'(j_{n})$ found in the A \& S p 409. 

Now write:

\begin{eqnarray}
P=\sum_{s}\frac{\theta}{\Theta}p_{s}(u)J_{1}(j_{s}\theta /\Theta),
\end{eqnarray}

and

\begin{eqnarray}
-\frac{1}{u^{2}-1}\beta\frac{d\beta}{dP}=\sum_{s}\frac{\theta}{\Theta}K_{s}(u)J_{1}(j_{s}\theta /\Theta).
\end{eqnarray}

Then multiply (43) by $J_{1}(j_{n}\theta /\Theta)$ and integrate from $0$ to $\Theta$; on division by $\frac{1}{2}[J_{1}'(j_{n})]^{2}$ we get, since $(\theta /\Theta) J_{1}(j_{s}\theta /\Theta )$ satisfies (67);

\begin{eqnarray}
\frac{d}{du}\Big[(u^{2}-1)\frac{dp_{n}(u)}{du}\Big]-\frac{j_{n}^{2}}{\Theta^{2}}p_{n}(u)=K_{n}(u).
\end{eqnarray}

The above equation is Legendre's with a known right hand side $K_{n}(u)$. This may be solved using the boundary conditions by variation of parameters. We write each $p_{n}(u)$ in the form

\begin{eqnarray}
p_{n}(u)=A_{n}(u)P_{\nu(n)}(u)+B_{n}(u)Q_{\nu(n)}(u),
\end{eqnarray}

where $\nu(n)(\nu(n)+1)=j_{n}^{2}/\Theta^{2}$ and $P_{\nu}$, $Q_{\nu}$ are Legendre functions satisfying

\begin{eqnarray}
\frac{d}{du}\Big[(1-u^{2})\frac{dP_{\nu}}{du}\Big]+\nu(\nu+1)P_{n}=0,
\end{eqnarray}

and $Q_{n}$ likewise. We choose $A_{n}$ and $B_{n}$ to satisfy

\begin{eqnarray}
A_{n}'(u)P_{\nu(n)}(u)+B_{n}'(u)Q_{\nu(n)}=0.
\end{eqnarray}

Thus

\begin{eqnarray}
p_{n}'=A_{n}P_{\nu}'(u)+B_{n}Q_{\nu}'(u).
\end{eqnarray}

Substituting (78) into (74) the terms with $A$ and $B$ undifferentiated vanish since $P_{\nu}$ and $Q_{\nu}$ satisfy Legendre's equation so the only terms that survive are

\begin{eqnarray}
(1-u^{2})[A_{n}\nu P_{\nu}'+B_{n}Q_{\nu}']=K_{n}(u). 
\end{eqnarray}

We solve (77) and (79) using the Wronskian $W=PQ'-Q'P$. But the Wronskian is $-(u^2-1)^{-1}$ see A \& S 8.1.9., so

\begin{eqnarray}
A_{n}(u)-A_{n}(u_{b})=\int_{u_{b}}^{u} K_{n}Q_{\nu(n)}du,
\end{eqnarray}

similarly,

\begin{eqnarray}
B_{n}=-\int_{u_{t}}^{u} K_{n}P_{\nu(n)}du,
\end{eqnarray}

where $u_{b}$ is the value of $u$ at the surface where the field expands at some given time, and $u_{t}$ is the value of $u$ at highest point of the cocoon $P+P_{0}$. $A_{n}(u_{b})$ can be determined by analysing $P(u_{b},\theta)$ into $J_{1}$ and by (75)

\begin{eqnarray}
A_{n}(u_{b})P_{\nu}(u_{b})+B_{n}(u_{b})Q_{\nu}(u_{b})=p_{n}(u_{b})
\end{eqnarray}

One must choose the end points of the integrations to ensure the boundary conditions are satisfied. Indeed formula (72) provides accurate results when the first few terms are evaluated and we are not very close to the extreme relativistic limit. In the paradigm solved for $l=0.1$ the deviation between the reconstructed solution and the solution satisfying (56), which holds inside the lobe where $P>P_{0}$, is $0.2\%$ at $u=1.5$ and $0.3\%$ at $u=1.05$, $0.8\%$ at $u=1.01$ and $4\%$ at $u=1.001$ by taking into account the first fifteen terms (Figures 1-3). As $u$ comes closer to unity the deviation between the reconstructed solution and the self-similar one increases. This deviation is systematic as the reconstructed solution always gives a flux function that is greater than the self-similar one. This is because the self-similar solution is proportional to $u^2-1$ and becomes $0$ at $u=1$, whereas the Legendre P functions converge to 1 at the top. Very close to the top of the lobe we reconstruct the solution by analysing $P(u_{t},\theta)$ into Bessel functions, then by virtue of (76) and setting $B_{n}(u_{t})=0$ because of (81) we can determine $A_{n}(u_{t})$. Then from (82) it is clear that for $1<u<u_{t}$ $A_{n}(u)=A_{n}(u_{t})$. Thus we have a complete description of the field everywhere.   

\begin{figure}
\begin{center} 
\epsfxsize=8.7cm 
\epsfbox{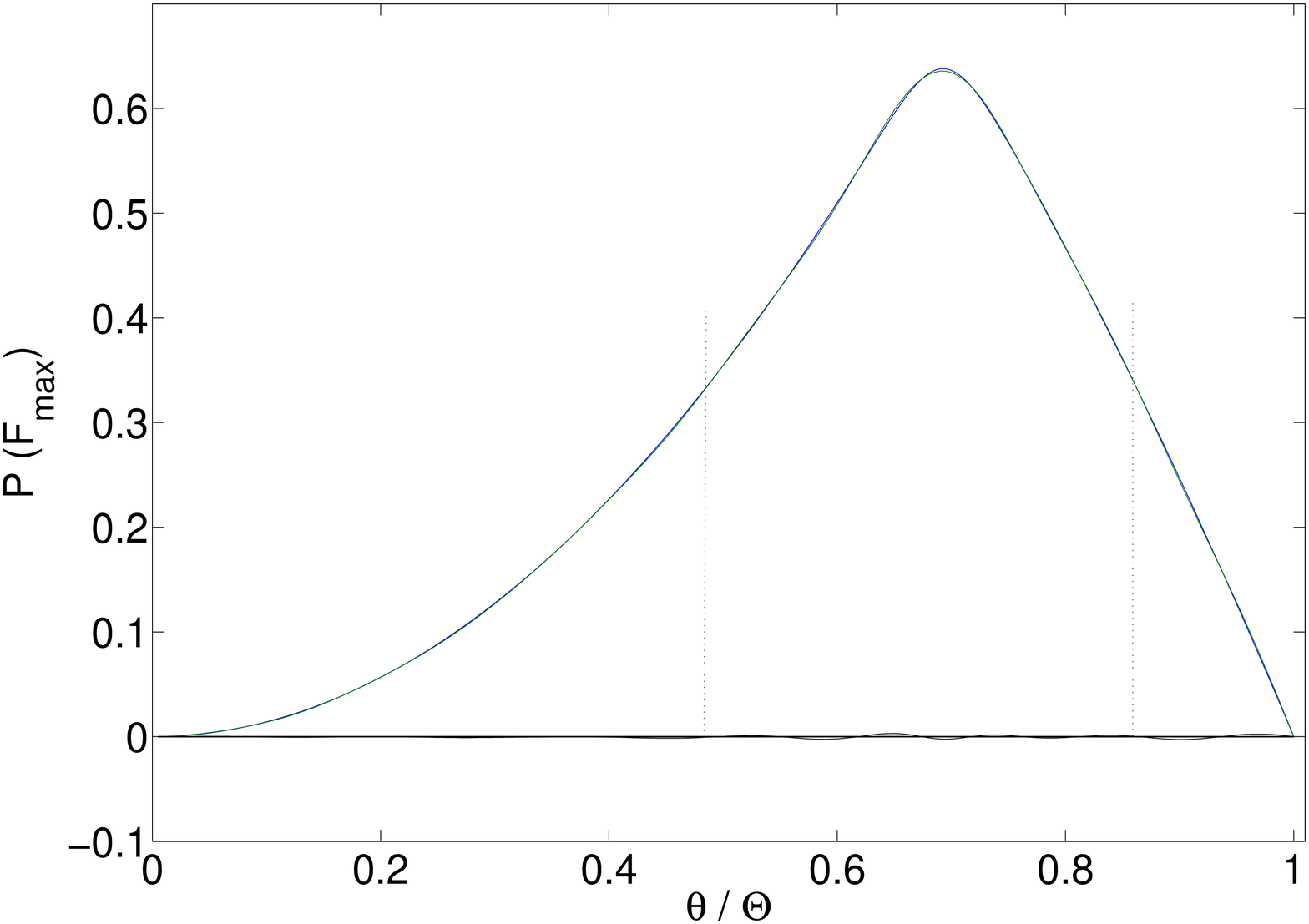}
\caption[]{The flux function $P$ for $u=1.5$ and $l=0.1$, given by equation (63) and reconstructed by using equation the first 15 components of (100). The small difference of the two forms is plotted at the bottom. The deviation of the two does not exceed $0.2\%$ of the maximum. The part of the plot between the vertical dotted lines corresponds to $P>P_{0}$.}
\label{}
\end{center}
\end{figure}

\begin{figure}
\begin{center} 
\epsfxsize=8.7cm 
\epsfbox{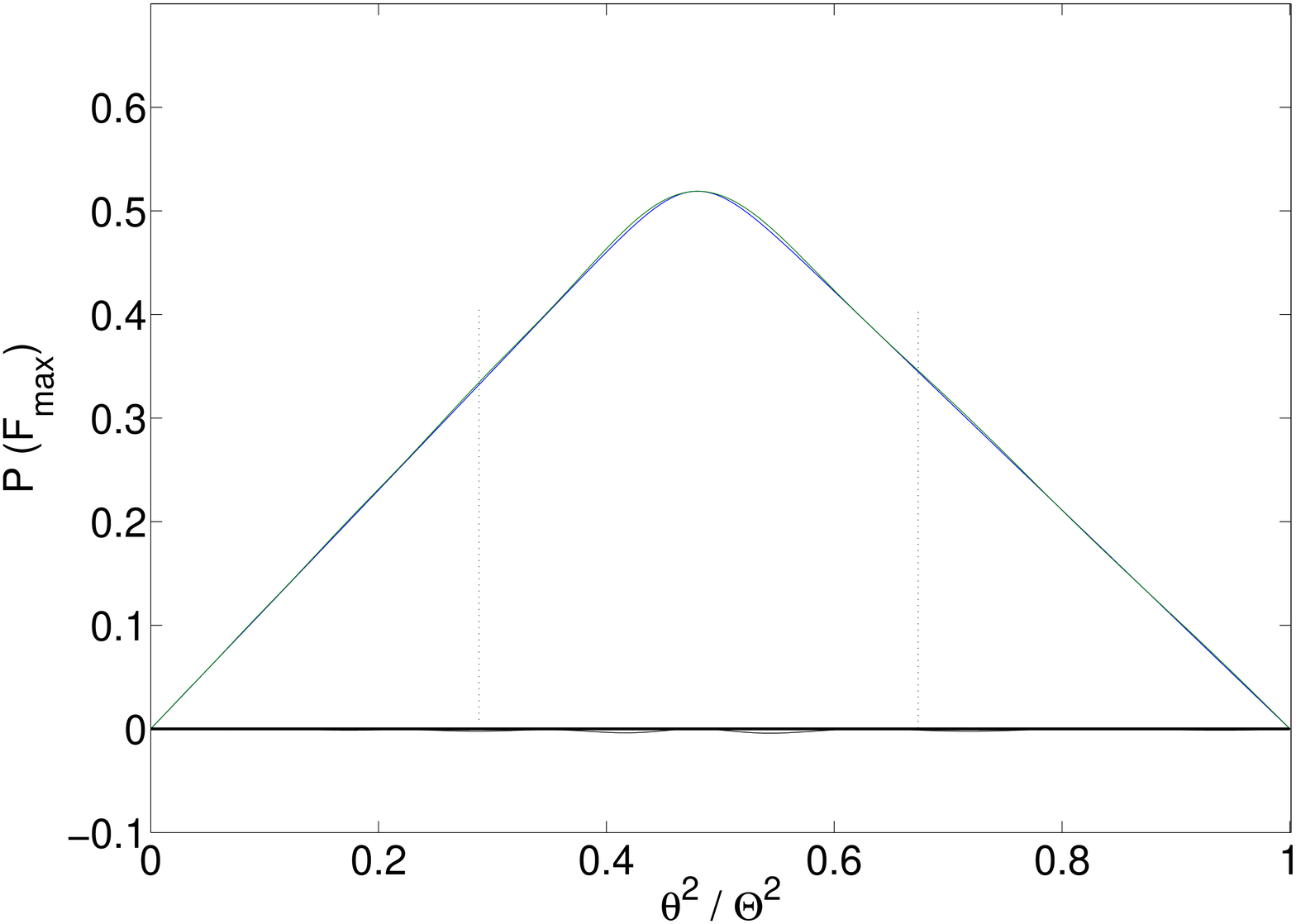}
\caption[]{The flux function $P$ for $u=1.01$ and $l=0.1$, given by equation (63) and reconstructed by using equation the first 15 components of (100). The difference of the two forms is plotted at the bottom. The deviation of the two does not exceed $0.8\%$ of the maximum. The part of the plot between the vertical dotted lines corresponds to $P>P_{0}$.}
\label{}
\end{center}
\end{figure}

\begin{figure}
\begin{center} 
\epsfxsize=8.7cm 
\epsfbox{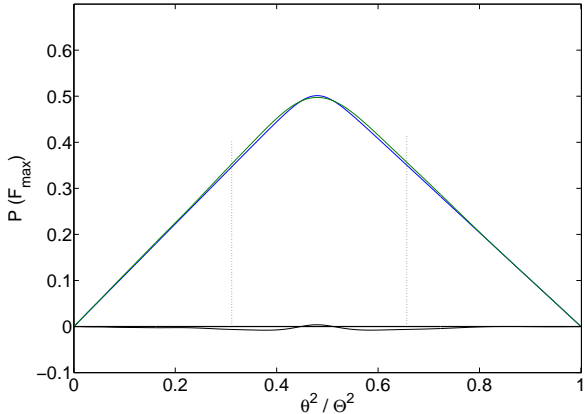}
\caption[]{The flux function $P$ for $u=1.005$ and $l=0.1$, given by equation (63) and reconstructed by using equation the first 15 components of (100). The difference of the two forms is plotted at the bottom. The deviation of the two does not exceed $1.3\%$ of the maximum. The part of the plot between the vertical dotted lines corresponds to $P>P_{0}$.}
\label{}
\end{center}
\end{figure}

\section{Physical quantities}

In this section we evaluate physical quantities such as the twist, the pressure and the energy corresponding to the mathematical solutions we describe above. We use quantities appropriate for $\gamma$-ray bursts of the long-soft type. We study two cases; one with semi-opening angle of $0.2 rad$ which is the maximum value for the assumption of narrow cones to hold and one with semi-opening angle of $0.03 rad$ which is closer to what is expected for $\gamma$-ray bursts. We find that in the latter case the fields are more twisted so the energy of the initial purely poloidal field is amplified by a greater factor.

\subsection{Twist}

The form of the field defines the distribution of the twist. For the reasons discussed in the section 3.2.1 we use the the solution for the flux function given by (63) for $P>P_{0}$. We have taken as a given that the magnetic field is contained within a cone. Although a detailed hydrodynamic study is needed in order to find the shape of the magnetic cavity; a pressure environment that depends in $r$ like $p\propto r^{-2l-4}$ leads to the conical jets.

Inspection of (62) shows that toroidal torque $\beta$ increases with $P$ when $P>P_{0}$ but it is zero for $P \leq P_{0}$. However our solution (63) shows that the lines of force for large $P$ do not go to large radii where the twist is. Thus the twist of the lines is greatest for some $P=P_{m}$ greater than $P_{0}$ and less than $F_{max}$. When $l$ is large $P_{m}$ is close to zero, but as $l$ decreases it shifts closer to $F_{max}$. The maximum twist increases as $l$ decreases (Figures 4 and 5). Smaller $l$ corresponds to fields that decrease slower with $r$ and at the same time they demand greater twist.

\begin{figure}
\begin{center} 
\epsfxsize=8.7cm 
\epsfbox{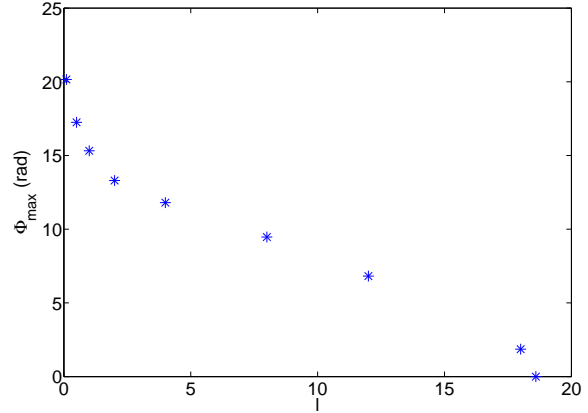}
\caption[]{The maximum twist with $l$ for a cone of semi-opening angle of 0.2 rad. As $l$ decreases the maximum twist inside the jet becomes greater. It becomes zero at $l=18.6$.}
\label{}
\end{center}
\end{figure}

\begin{figure}
\begin{center} 
\epsfxsize=8.7cm 
\epsfbox{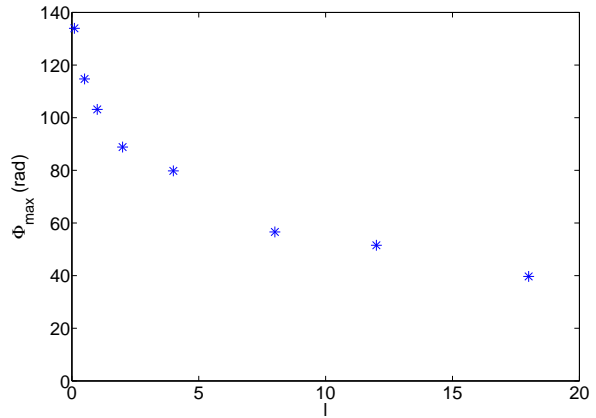}
\caption[]{The maximum twist with $l$ for a cone of semi-opening angle of 0.03 rad. As $l$ decreases the maximum twist inside the jet becomes greater. Compared to figure 4 the twist is bigger, it drops to zero for $l=127.2$ which is not included in this plot. In general the qualitative behaviour of $\Phi_{max}(l)$ in similar to this of the wider cone.}
\label{}
\end{center}
\end{figure}

\begin{figure}
\begin{center} 
\epsfxsize=8.7cm 
\epsfbox{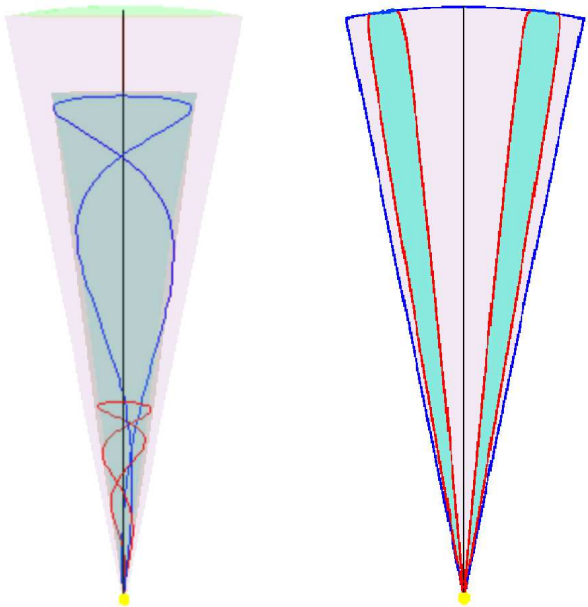}
\caption[]{Left: Field lines in the case of $l=0.1$ in the jet of semi-opening angle $0.2rad$. The field lines correspond to fluxes 0.6 red and 0.7 blue of the total flux. The outer pink cone (light coloured in black and white) is the cone confining the jet, whereas the blue surface (dark coloured in black and white) is the surface corresponding to flux 0.6 on which the blue field line lies.\\
Right: The lobe and the cocoon for the same jet. The lobe in blue comes very close to the top of the jet. The size of the cone in both cases is $R_{*}$. Narrower jets have many more twists.}
\label{}
\end{center}
\end{figure}

The lines of force of the magnetic field are given:

\begin{eqnarray}
\frac{dr}{B_{r}}=\frac{rd\theta}{B_{\theta}}=\frac{r\sin\theta d\phi}{B_{\phi}}.
\end{eqnarray}

In the region where $P<P_{0}$ there is no twist in the field, as $\beta=0$, thus $B_{\phi}=0$. In the region where $P>P_{0}$ the field can be evaluated by virtue of (63) and (8) and by using the second equality of (83) the total twist $\Phi_{P}$ for a field line determined by $P$ is given by:

\begin{eqnarray}
\Phi_{P}=\int_{0}^{\Phi}d\phi=\frac{\beta(P)}{lP}\int_{\theta_{1}}^{\theta_{2}}\frac{d\theta}{u\sin \theta},
\end{eqnarray}

where $\theta_{1}$ and $\theta_{2}$ are the polar angles of the points where the field line rises and sinks at the base. We then express $u$ as a function of $\theta$ for some given $P$ and we integrate. This gives the detailed shape of the field lines (Figure 7). Most of the twist is concentrated near the top of each field line. 

From integral (84) for $\theta_{1}$ and $\theta_{2}$ corresponding to the footpoints of the field lines we evaluate the total twist a field line carries. The maximum twist depends on the opening angle of the jet. The overall behaviour of $\Phi_{max}$ can be approximated by the analytical formula for the total twist in the non-relativistic case \citep{2008MNRAS...385..875G} $\frac{\sqrt{2}\pi}{\sin\Theta}$ (Figure 7) or  $\frac{\sqrt{2}\pi}{\Theta}$ for the case of narrow cones we are interested in this study. Thus  the smaller the opening angle the more twisted the field is, giving evidence for the relation between collimation and the number of the turns performed. Although the total twist is inversely proportional to the semi-opening angle, the constant of proportionality is somewhat smaller than what was found from the non-relativistic analytical formula.

The total twist for a given field line does not depend on time, indeed they merely expand with the expanding frame. Nevertheless the field lines rotate as they expand relative to the stationary frame. The azimuthal velocity component given in equation (17) describes that rotation of the field lines and is a fraction of the speed of light, its maximum value is $0.6c$. As we have a non-stationary solution of Maxwell's equations, problems at the light cylinder cannot occur, and even a radius for the light cylinder is not defined. Rotation has been observed in non-relativistic jets emerging from T-Tauri stars \citep{2007ApJ...663..350C}.

\begin{figure}
\begin{center} 
\epsfxsize=8.7cm 
\epsfbox{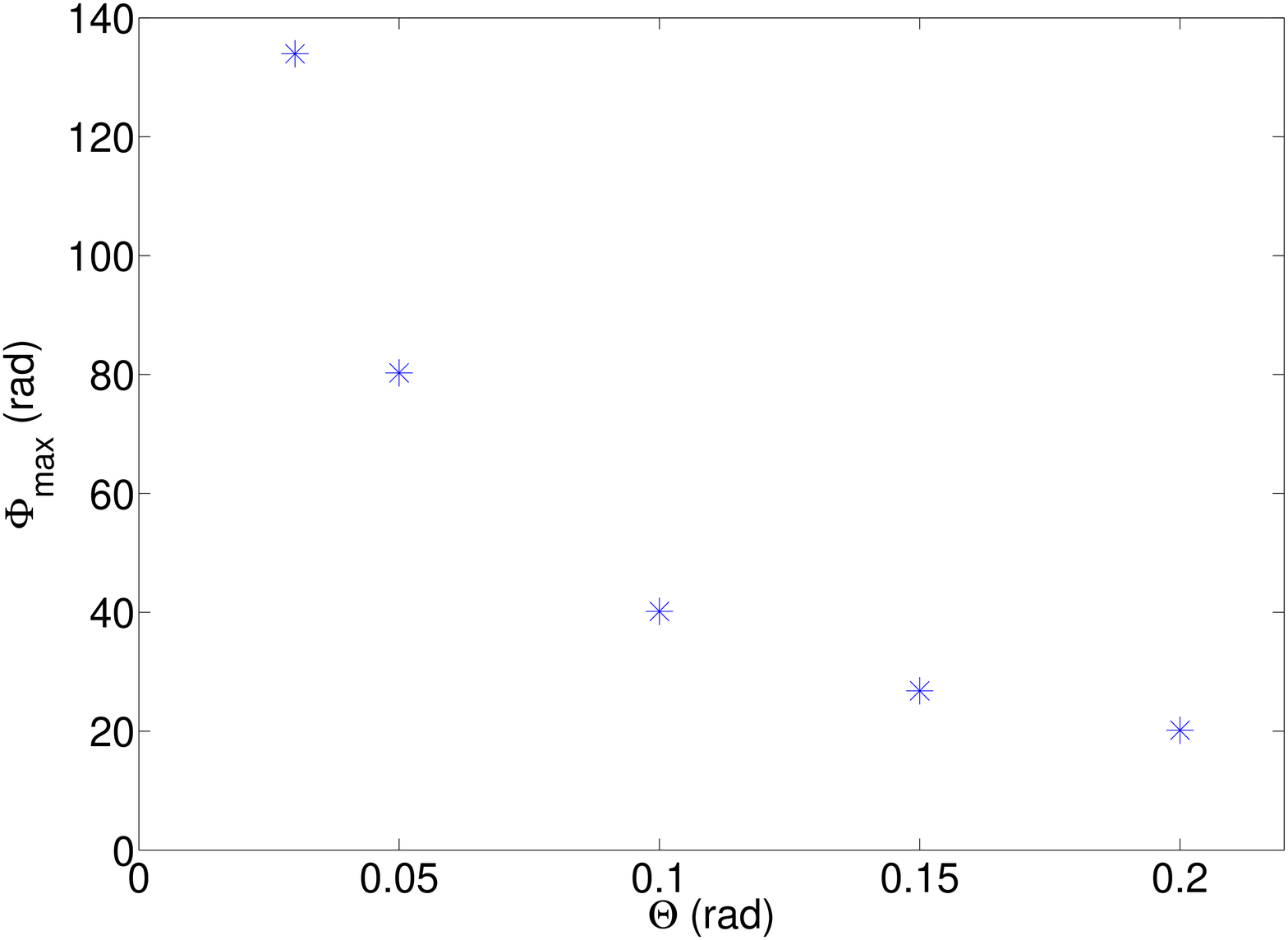}
\caption[]{The maximum twist as a function of the semi-opening angle for $l\to 0$. The maximum twist is inversely proportional to the semi-opening angle, but instead of proportionality constant $\sqrt{2}\pi =4.40$ it is $\Phi_{max}=\frac{4.03}{\Theta}$ }
\label{}
\end{center}
\end{figure}

\subsection{Energy}  

We can easily find the pressure and the energy carried by a given electromagnetic field since we know the actual fields. As the energy density of an electromagnetic field is proportional to the pressure we are going to treat them together. The energy density is given by

\begin{eqnarray}
e=\frac{1}{8\pi}(\mathbf{B}^{2}+\mathbf{E}^{2}).
\end{eqnarray}
 
\subsubsection{Energy distribution}
From equations (8) and (9) and the flux function $P$ from equation (63) we can evaluate the pressure at some given time $t$ (Figures 8 and 9). The energy is mainly concentrated near the base of the jet and near the top of the jet. It is expected that the energy is strong near the base as the fields decrease with distance from the centre. However unlike other cases studied where the fields extend to infinity, there is a very strong field near the top. This is because of the turn over of the field lines, which are confined inside the light sphere, so they have to close. Thus, they have a very strong $B_{\theta}$ component. 

\begin{figure}
\begin{center} 
\epsfxsize=8.7cm 
\epsfbox{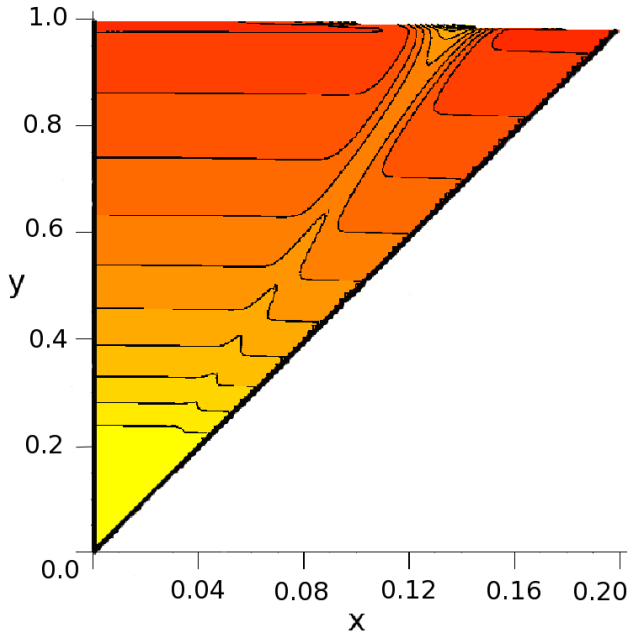}
\caption[]{The energy density and the static pressure in a section of the jet. The contour lines correspond to surfaces of constant pressure and differ with the consecutive by a factor of two. They are stronger at the bottom and they have a hot spot near the top. This is because the field lines turn back near the top giving a strong $B_{\theta}$ component in the field. They increase as we move towards the centre of the jet reaching a maximum and then decrease in the outer part because of the strong $B_{\phi}$ field inside the lobe. The pressure near the axis is stronger than the pressure near the edge because of hoop stresses. The axes are not to scale, the units of the axes are fraction of the total jet length at that time.}
\label{}
\end{center}
\end{figure}

\begin{figure}
\begin{center} 
\epsfxsize=8.7cm 
\epsfbox{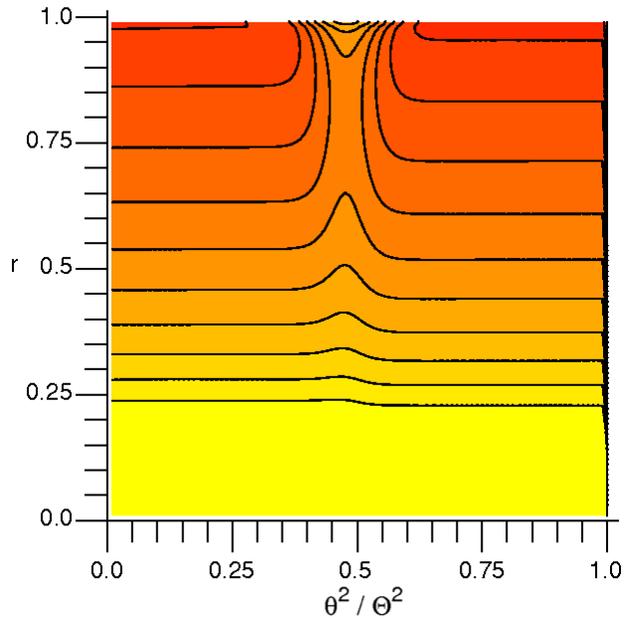}
\caption[]{The energy density and the pressure in $\theta^2$ and $r$. The same conclusions as from figure 8, but in this case one can see the symmetry of the structure in $\theta^{2}$.}
\label{}
\end{center}
\end{figure}

As we approach the light horizon where the expansion velocity is equal to the speed of light the energy strongly depends on the smoothing of the fields. If we considered an infinitely fast turning on of the magnetic field then it would carry infinite energy, however this is  unphysical as it takes finite time to switch on. That time scale determines the energy of the initial pulse; it is inversely proportional to the time-scale of switching on the dipole. The physical system we have in mind is the following: as the core of a massive star collapses to form a compact object or a black hole it spins much faster and winds up the magnetic field where it connects to the rest of the star or to any disc that may form. This leads to a dramatic increase in the energy by two orders of magnitude in the case of a narrow jet, as shown in the next section. All the twist is imposed in this stage. The concentrated energy of the field leads to an explosion. The magnetic field expands in the star and a jet forms. The scales chosen for this study are $R_{*}=10^{13}cm$ for the size of the jet. This is the typical size of the jet related to the prompt emission of a $\gamma$-ray burst. $R_{c}=10^{7}cm$ for the smoothing scale that is a few times the length scale of the Schwarzschild radius of a black hole of $10 M_{\sun}$.

We consider a magnetic field of order $B_{0}=10^{13}-10^{14}Gauss$ at $R_{0}=10^{9}cm$ powering the jet. This magnetic field comes from the central compact object, a discussion on the origin of strong magnetic fields in neutron stars can be found in \cite{2008AIPC..983..391S} and references therein. The strongest magnetic fields estimated for magnetars are $10^{15}Gauss$. We assume that such a field originates at $R_{c}$ and then expands outwards, as the surrounding pressure near the centre drops relatively slowly.  A constant pressure environment leads to a field which is confined within a cylinder without decreasing at all, we take an intermediate case for which the field between $R_{c}$ and $R_{0}$ drops by two orders of magnitude. Above this radius and below $R_{*}$ we apply the self-similar solutions inside a narrow cone. These give a total energy of order $10^{51}erg$ in the jet from which $10^{49}erg$ is due to the switching on of the field concentrated in a very thin layer just below the light sphere. The width of this layer is $10^{7}cm$ and the fields contained are of order $10^{9}Gauss$ whereas the field at the main body of the jet at $5\times 10^{12}cm$ is of order of $10^{5}Gauss$. The energies quoted above and on the tables are at $t=300s$ after the initial explosion in the centre of the star, when the top of the jet reaches the surface of the star.
 
\begin{table}
 \centering
 \begin{displaymath}
% use packages: array
\begin{array}{|r|r|r|r|r|}
\hline
l&\Phi &E_{Pol}            &E_{Tor}            & E_{Tot}  \\
 &(rad)     &(\times 10^{51}erg)&(\times 10^{51}erg)&(\times 10^{51}erg) \\
\hline
0.1 & 20.16 & 0.916 & 0.083 & 1.000 \\
0.5 & 17.25 & 0.601 & 0.165 & 0.770 \\
1.0 & 15.32 & 0.426 & 0.244 & 0.671 \\
2.0 & 13.30 & 0.271 & 0.224 & 0.496 \\
4.0 & 11.80 & 0.163 & 0.179 & 0.342 \\
8.0 & 9.47  & 0.098 & 0.127 & 0.227 \\
12.0 & 6.82 & 0.082 & 0.086 & 0.168 \\
18.0 & 1.86 & 0.075 & 0.010 & 0.085 \\
18.6 & 0.00 & 0.075 & 0.000 & 0.075 \\

\hline
 \end{array}
 \end{displaymath}

 \caption{The maximum twist and the energy of the various components of the magnetic field for a given flux $F_{max}=2\times 10^{30}Gauss \times cm^2$, corresponding to an average magnetic field of order $10^{13} Gauss$ emerging from a spherical cup of opening semi-angle $0.2rad$ and radius $10^{9}cm$. As $l$ decreases the field gets twisted and its energy amplified.}
 \label{Table 1}
\end{table}

\begin{table}
 \centering
 \begin{displaymath}
% use packages: array
\begin{array}{|r|r|r|r|r|}
\hline
l&\Phi&E_{Pol}            &E_{Tor}            & E_{Tot}  \\
 &(rad)     &(\times 10^{51}erg)&(\times 10^{51}erg)&(\times 10^{51}erg) \\
\hline
0.1 & 133.96 & 0.916 & 0.085 & 1.000 \\
0.5 & 114.72 & 0.600 & 0.205 & 0.805 \\
1.0 & 103.15 & 0.423 & 0.217 & 0.640 \\
2.0 & 88.85  & 0.269 & 0.186 & 0.455 \\
4.0 & 77.80  & 0.155 & 0.133 & 0.288 \\
8.0 & 60.10  & 0.085 & 0.083 & 0.168 \\
12.0 & 51.53 & 0.059 & 0.061 & 0.119 \\
18.0 & 39.70 & 0.040 & 0.045 & 0.085 \\
127.2 & 0.0  & 0.006 & 0.000 & 0.006 \\

\hline
 \end{array}
 \end{displaymath}

 \caption{The maximum twist and the energy of the various components of the magnetic field for a given flux $F_{max}=10^{30} Gauss \times cm^2$, corresponding to an average magnetic field of order $10^{14} Gauss$ emerging from a spherical cup of opening semi-angle $0.03rad$ and radius $10^{9}cm$. As $l$ decreases the field gets twisted and its energy amplified. This case clearly illustrates that a jet of smaller opening angle can be more twisted than one of a bigger opening angle. The $l$ for which the field does not have any toroidal component is $127.2$. The energy of the purely poloidal structure is much smaller than this of the one that is twisted.}
 \label{Table 1}
\end{table}

\subsubsection{Energy as a function of twist}

The total energy depends on $l$ (Figures 10 and 11). It can be seen (tables 1 and 2; figures 4 and 5) that $l$ parametrises the twist. We prefer the use of the twist as parameter as it describes a physical process rather than $l$ which is just an index. Thus a decrease in $l$ can be viewed as an increase on the differential rotation at the base. When differential rotation is imposed at the base there is a toroidal component in the field which increases the energy of the configuration, and in addition to that the poloidal part of the field expands to reach a force free state. This dual process leads to an increase in the total energy. The factor by which the total energy is amplified depends on the maximum twist the configuration can tolerate and thus the opening angle which determines the maximum twist. Narrower cones get a greater amplification on their total energy; in the limiting case which is equivalent to an infinite cylinder, the twist has no constraint, thus the energy can be infinitely amplified.

\begin{figure}
\begin{center} 
\epsfxsize=8.7cm 
\epsfbox{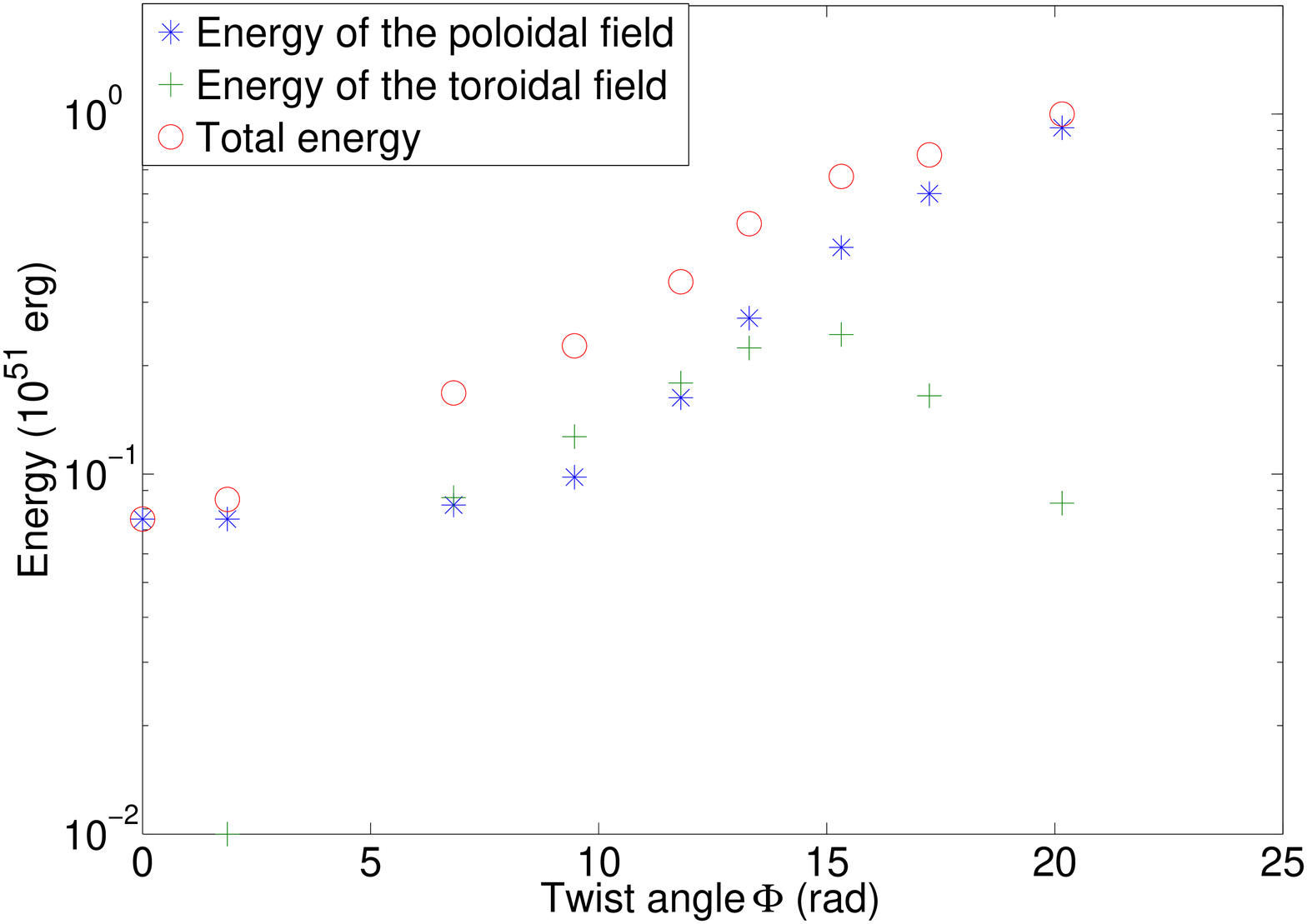}
\caption[]{The energy as a function of the twist for a cone of semi opening angle of $0.2 rad$. It is clear that as the twist increases energy is injected in the system. Initially all the the energy is in the poloidal field, however the energy of the toroidal field, increases reaches a maximum and then decreases. Unlike other cases of spherical symmetry where the field lines are allowed to reach infinity there is some contribution of the toroidal magnetic field to the total energy of the field when it reaches its maximum twist.}
\label{}
\end{center}
\end{figure}

\begin{figure}
\begin{center} 
\epsfxsize=8.7cm 
\epsfbox{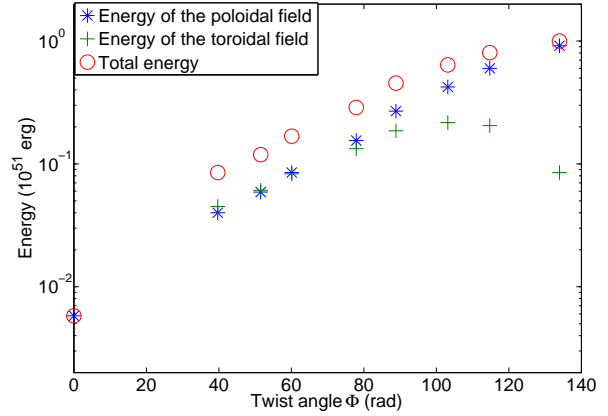}
\caption[]{The energy as a function of the maximum twist for a cone of semi opening angle of $0.03 rad$. The conclusion are similar to those of figure 10. The qualitative behaviour of energy with twist is the same for smaller opening angles.}
\label{}
\end{center}
\end{figure}

\subsubsection{Energy flow}

As the whole structure expands there is a flow of energy in the jet. The energy contained in an element of the the jet from angle $\theta$ to $\theta+d\theta$ and in velocity space between $v$ and $v+dv$ is

\begin{eqnarray}
E_{1}=\frac{1}{16 t \pi^{2} v^{2} \sin\theta} \Big(\big[\big(\frac{\partial P}{\partial v}\big)^{2}+\big[v^{2}\big(\frac{\partial P}{\partial v}\big)^{2}+T^{2}\big](1+v^{2})\Big)d\theta dv.
\end{eqnarray}

This energy is inversely proportional to time. The Poynting vector for the fields is

\begin{eqnarray}
\mathbf{S}=\frac{c \mathbf{E} \times \mathbf{B}}{4 \pi},
\end{eqnarray}

substituting from (8) and (9) 

\begin{eqnarray}
\mathbf{S}=\frac{c}{16 \pi^{2} r^{4}}\Big(\frac{v}{1-\mu^{2}}\Big(T^{2}+v^{2}\Big(\frac{\partial P}{\partial v}\Big)^{2}\big)\mathbf{\hat{r}} \nonumber \\ 
- \frac{v^{2}}{\sqrt{1-\mu^{2}}}\frac{\partial P}{\partial v} \frac{\partial P}{\partial \mu} \bm{\hat{\theta}} +\frac{vT}{\sqrt{1-\mu^{2}}}\frac{\partial P}{\partial \mu} \bm{\hat{\phi}}\Big). & &
\end{eqnarray}

The $\theta$ component of the Poynting vector is zero at the boundaries of the jet and on the axis, so it is consistent with the demand that the jet is constrained within a cone and there is no flow of energy out of the sides of the cone. We now apply the Poynting theorem for the expanding volume

\begin{eqnarray}
-\frac{d}{cdt}\int\frac{E^{2}+B^{2}}{8 \pi} dV= \nonumber \\
=- \int \frac{E^{2}+B^{2}}{8 \pi} \mathbf{v}\cdot d\mathbf{S} +\int \frac{\mathbf{E} \times \mathbf{B}}{4\pi}\cdot d \mathbf{S}.
\end{eqnarray}

As opposed to the usual form for the Poynting theorem, there is an extra term in the right hand side because of the expansion of the volume. The term in the left hand side is the integral of equation (86). The terms in the right hand side show in which direction the energy is transfered in $v$ space. The integrals should be performed on the surfaces of the inner and the outer boundary of the expanding volume. Nevertheless, by integrating only on one surface we can find whether the energy flow is towards larger or smaller $v$.

\begin{eqnarray}
\int \frac{\mathbf{E} \times \mathbf{B}}{4 \pi}d \mathbf{S}-\int \frac{E^{2}+B^{2}}{8 \pi}\mathbf{v} \cdot d\mathbf{S}= \nonumber \\
=\frac{1}{8 \pi^{2} v t^{2}}\int \Big(\big[T^{2}+v^{2}\big(\frac{\partial P}{\partial v}\big)^{2}\big]\frac{1-v^{2}}{2}-\frac{1}{2}\big(\frac{\partial P}{\partial \theta}\big)^{2}\Big) \frac{d \theta}{\sin \theta}.
\end{eqnarray} 

The integral of equation (90) is negative when integrated for the solutions of $P$ and $T$ found in section 3.2. Thus the energy flows towards smaller $v$. Nevertheless the energy flow relative to fixed axes is outwards everywhere. This discrepancy occurs because the energy flow as described by the Poynting vector is slower than the velocity due to the uniform expansion.

The region at the top has to be treated separately. Near the top $T=0$ and it is only the component due to $\frac{\partial P}{\partial v}$ that contributes to the energy. The energy in the top is compressed in a thin shell (region II). The width of this shell is $R_{c}$. Thus near the top the flux function drops from $P(v=1,\theta)$ to $0$ within a distance $R_{c}$ as described in Appendix B for the smoothing of the fields in the case of self-similar multipoles. $P(v=1,\theta)$ is approximated with good accuracy by $P_{0}f(\theta)$, where $P_{0}$ is defined for equation (62) and $f(\theta)$ is the solution to equation (56), note that the exact value of $P(v=1,\theta)$ is not given by equation (63) but it is a sum of multiples of Bessel and Legendre functions. After we smooth the field we find that the energy contained in region II

\begin{eqnarray}
E_{top}=\frac{1}{4 \pi^{2}}\frac{P_{0}^{2}}{R_{c}}\int_{0}^{\Theta}\frac{f(\theta)^{2}}{\sin \theta} d\theta.
\end{eqnarray}

Thus the energy of the shell (region II) is constant.

\subsection{Pressure}

\subsubsection{Static pressure}
As the pressure of the magnetic field is proportional to $r^{-4-2l}$ for the case of the self-similar solutions in the non-relativistic region the field shall demand a special pressure environment so that it is contained within a cone. Let us assume that the pressure of the surrounding medium is described by a power-law $r^{-\zeta}$, thus as the field is differentially rotated it will expand primarily up the axis and if the index of the pressure of the medium ${\zeta}$ is greater that $4-2l$ then the jet will not have a constant opening angle but will expand sideways. 

\subsubsection{Stagnation pressure}

Near the top the case is rather different, as there is a particularly strong ram pressure. This is because the head of the jet encounters the gas of the medium in a velocity $v_{h}$ that is close to the speed of light, let this velocity correspond to some Lorentz factor $\gamma$. As there is only a $B_{\theta}$ component there, in the comoving frame the magnetic field is:

\begin{eqnarray}
B_{\theta}'=\frac{1}{\gamma}B_{\theta}.
\end{eqnarray}

This magnetic field encounters an environment of density

\begin{eqnarray}
\rho'=\gamma \rho,
\end{eqnarray}

where $\rho$ is the density of the medium in its rest frame. The $\gamma$ factor is because the coordinate in the direction of expansion is shrunk by a factor of $\gamma$. Balance of momenta gives a stagnation pressure in the comoving frame

\begin{eqnarray}
p_{s}'=\gamma^{2} \rho v_{h}^{2}=\frac{B^{'2}_{\theta}}{8 \pi}=\gamma^{-2}\frac{B_{\theta}^{2}}{8 \pi}.
\end{eqnarray}

Setting the stagnation pressure equal to the magnetic pressure of the top field we can determine the $\gamma$ that the magnetic field reaches as it expands. This depends on the magnetic field and the density the jet encounters; as $B_{\theta}$ and $\rho$ are known we can easily find the value of $\gamma$. For densities around $10^{-5}g/cm^{3}$ a Lorentz factor of $2$ is found,whereas for densities smaller than $10^{-8}g/cm^{3}$ Lorentz factors greater than 10 are found, and for the case of an HII region surrounding a massive star the jet becomes extremely relativistic with $\gamma>>100$. However, near the centre of the star the density greatly exceeds the value of $10^{-5}gr/cm^{3}$, thus the jet pushes its way through the material at non-relativistic speeds. However it is possible that a funnel forms by an other mechanism. In the case of $\gamma$-ray bursts there is a black hole formed in the centre of a rotating star thus, the material on the axis will fall into the black-hole. Then there will be an underdense region on the axis allowing the expansion of the jet, see e.g. \cite{1978PhyS...17..185L}.

\section{Applications}

Relativistic jets were first found emanating from radio galaxies and quasars, then more spectacularly still from $\gamma$-ray burst sources and microquasars. Related but non-relativistic jets arise from young stars with accretion discs and generate Herbig-Haro objects. We consider that all these phenomena come from the magnetic energy generated when a magnetic field is wound up by the differential rotation of its footpoints. The opening angles of many of these jets is of order $4^{o}$, giving semi angles of $2^{o}$ or so. 

It is believed that massive Wolf-Rayet stars in rapid rotation generate the long-soft $\gamma$-ray bursts when their cores collapse eventually to make black holes and the associated supernovae \citep{2003ApJ...591..288H}. Then the differential rotation is between the collapsing core of the star and the enclosing envelope. It has also been suggested that the outcome of the collapse is a strongly magnetised compact object \citep{1992ApJ...392L...9D} releasing energy in magnetic field. This magnetic energy is generated by differential rotation deep within the star. While the strong magnetic field expands to balance the external gas pressure the density of the external envelope initially prevents relativistic expansion along the rotation axis. Under expansion at constant flux the field in a magnetic bubble or cavity varies, with length scale, $L$, as $L^{-2}$ and the magnetic pressure as $L^{-4}$. Thus while it is within the body of the star the magnetic cavity is collimated when the ambient pressure falls less rapidly than $r^{-4}$ subtending smaller angles at the star's centre as $r$ increases. Such configurations of non-relativistic jets were described in \cite{2007MNRAS.378..409S} and references therein and have been applied to $\gamma$-ray bursts by \cite{2006ApJ...647.1192U}. In the outer parts of the star the pressure falls more rapidly than $r^{-4}$ and the head of the magnetic cavity accelerates until it breaks through the stellar surface and expands relativistically. Numerical studies of ultrarelativistic MHD jets have been done by \cite{2004ApJ...608..378F}, where it was found that high collimation angles can be achieved by a configuration where the toroidal component of the magnetic field dominates. When the central object is a rotating black hole then the Kerr geometry was taken into account \citep{1997A&A...319.1025F}. In the case we study the main source of confinement is ram pressure on the low density stellar wind. 

\section{Conclusions}
In this paper we have solved analytically the equation proposed by Prendergast for relativistically expanding axisymmetric self-similar force-free fields. This equation is a relativistically moving extension of the Grad-Shafranov equation for static force-free fields with axial symmetry. 

To excellent accuracy our solution is given everywhere by equations (62) and (63) with the $f(\mu)$ function given by equation (57) or more accurately (60). The resulting electromagnetic fields are given by equations (8) and (9). All solutions of the Prendergast equation need to be averaged over a small time $\Delta \tau$ that corresponds to the light crossing time across the magnetic explosion. His equation assumes a point explosion. Exact solutions of Prendergast's equation give singular fields where the expansion speed is that of light but we have shown how averaging over $\Delta \tau$ yields genuine solutions of Maxwell's equations without singularities. Nevertheless this $\Delta \tau$ may be quite short and the resulting fields close to $r=ct$ are strong, carrying around $1\%$ of the system's energy. This may be related to $\gamma$-ray burst precursors. However, the fields at the top require further modification as the ram pressure of their relativistic motion is retarded by the interstellar medium. Considerations as those of \citep{2005ApJ...628..847R} are clearly important there, whatever drives the shock waves and it is there that the eponymous $\gamma$-rays themselves are generated. 

While this paper has given a basic mechanism that produces highly relativistic motion there is a big gap between that and the phenomenology of $\gamma$-ray bursts.

On the way to understanding the solutions of the fully relativistic Prendergast equation we were led to give purely poloidal solutions of Maxwell's equations for time-dependent multipoles, interesting in their own right. We have also seen how in the non-relativistic limit our problem is related to the solutions found in \cite{2006MNRAS.369.1167L} and \cite{1994MNRAS...267..146L}. 

Finally we have seen how small opening angles of the cone correspond to greater coiling of the magnetic field and thus to mush greater concentration  of energy into the jet.  

Setting some values for the magnetic field quoted from the stronger magnetars we find that we can power jets up to the energies of $\gamma$-ray bursts, for which the isotropic energy is between $10^{51}$ to $10^{54}$ erg \citep{2005RvMP...76.1143P}. In the case studied we found that the energy inside the jet is of order $10^{51} erg$, where the isotropic energy is two orders of magnitude greater. The very strong magnetic field found at the top of the magnetic configuration can oppose the strong ram pressure because of expansion and swipe material out of the way for the jet to expand. This feature may be related to the precursor observed in some $\gamma$-ray bursts e.g. \cite{1991Natur.350..592M}, as it is the first to break out and carries a small fraction of the energy of the $\gamma$-ray burst. 

Jets of narrow opening angles can be more twisted than those of wide opening angles, giving evidence that collimation increases with twist. 

Finally a strong stagnation pressure is found at the head of the jet. It prevents the jet from being relativistic near the base but as it expands outwards and encounters underdense material it becomes extremely relativistic as is indeed observed in $\gamma$-ray bursts.

\section*{Acknowledgements}

We thank Uden Sherpa for his organisation of the Himalayan camp where the partial solution (63) was obtained and Profs Gough and Pringle for mathematical discussion of some of the finer points of the averaging of the singular fields in region (II). We also thank Drs Eldridge and Tout for the discussion of $\gamma$-ray burst progenitors. We thank the referee for demanding that we clarify the physics of rotation and energy flow in this solution.

{}

\appendix
\section{Dipole}

Following the formalism of \cite{1952} we express the electric and the magnetic fields in the following way, dot denotes differentiation with respect to time,
\begin{eqnarray}
\mathbf{E}=-\nabla \times \dot{\mathbf{\Pi}},
\end{eqnarray}

\begin{eqnarray}
\mathbf{B}=-\ddot{\mathbf{\Pi}}+\nabla(\nabla \cdot \mathbf{\Pi}).
\end{eqnarray}

The fields must satisfy Maxwell's equations in vacuum. By construction they satisfy 

\begin{eqnarray}
\nabla \cdot \mathbf{E}=0,
\end{eqnarray} 

\begin{eqnarray}
\nabla \times \mathbf{B}=\dot{\mathbf{E}}
\end{eqnarray}
 
The other two equations will let us determine function $\mathbf{\Pi}$. Substituting into $\nabla \cdot \mathbf{B}=0$ we take

\begin{eqnarray}
\nabla \cdot (\ddot{\mathbf{\Pi}}-\nabla^{2} \mathbf{\Pi})=0,
\end{eqnarray}

and by substituting in the equation $\nabla \times \mathbf{E}=-\dot{\mathbf{B}}$ we have:

\begin{eqnarray}
(\ddot{\mathbf{\Pi}}-\nabla^{2} \mathbf{\Pi})^{\cdot}=0.
\end{eqnarray}
  
Given that the fields that have physical meaning are expressed in forms of derivatives of $\mathbf{\Pi}$ we can choose it to satisfy the wave equation without overconstraining the fields. The proof that follows uses a gauge transformation of $\mathbf{\Pi}$.

\begin{eqnarray}
\widetilde{\mathbf{\Pi}}=\mathbf{\Pi}+ \nabla \times \mathbf{C},
\end{eqnarray}

for $\mathbf{\dot{C}}=0$. Indeed:

\begin{eqnarray}
\widetilde{\mathbf{E}}=-\nabla \times \dot{\widetilde{\mathbf{\Pi}}}= -\nabla \times \mathbf{\dot{\Pi}}=\mathbf{E},
\end{eqnarray}

\begin{eqnarray}
\widetilde{\mathbf{B}}=-\mathbf{\ddot{\widetilde{\Pi}}}+\nabla(\nabla \cdot \mathbf{\widetilde{\Pi}})=\mathbf{B}.
\end{eqnarray}

Integrating equations (A5) and (A6), we take:

\begin{eqnarray}
\mathbf{\ddot{\Pi}}-\nabla^{2}\mathbf{\Pi}= \mathbf{A},
\end{eqnarray} 

where $\nabla \cdot \mathbf{A}=0$ and $\dot{\mathbf{A}}=0$. Using the gauge transformation that leaves the fields unaffected as we have shown above, we can substitute in favour of $\mathbf{\widetilde{\Pi}}$, and equation (A10) becomes:

\begin{eqnarray}
\mathbf{\ddot{\Pi}}-\nabla^{2}\mathbf{\Pi}= \mathbf{A}-\nabla^{2}(\nabla \times \mathbf{C}).
\end{eqnarray}  

Note that the divergence of the RHS vanishes. Thus by choosing

\begin{eqnarray}
\nabla \times \mathbf{C}=-\frac{1}{4\pi}\int\frac{\mathbf{A}(\mathbf{r'})}{|\mathbf{r}-\mathbf{r'}|}d^{3}r',
\end{eqnarray}

we can make the RHS of (A11) vanish and therefore, equations (A5) and (A6) reduce to the wave equation.

The next task is to solve the wave equation. In axial symmetry with $\mathbf{\Pi}$ along $\mathbf{\hat{z}}$ everywhere, the wave equation becomes:

\begin{eqnarray}
\Big[\frac{1}{c^{2}}\frac{\partial^{2}}{\partial t^{2}}-\frac{1}{r^{2}}\frac{\partial}{\partial r}\Big(r^{2}\frac{\partial}{\partial r}\Big)-\frac{1}{r^{2}\sin \theta}\frac{\partial}{\partial \theta}\Big(\sin \theta \frac{\partial}{\partial \theta}\Big)\Big]\Pi=0,
\end{eqnarray} 

when $\Pi$, the magnitude of $\mathbf{\Pi}$, has spherical symmetry, the solution to the above equation describes a wave emerging from the centre and expanding outwards. It is 

\begin{eqnarray}
\mathbf{\Pi}=\frac{D(ct-r)}{r}\mathbf{\hat{z}},
\end{eqnarray} 

where $D(ct-r)$ is any function.

\section{Self-similar multipoles}

It is possible to extend the idea of the Hertzian dipole to describe a multipole. These multipoles have electric and magnetic fields confined between co-axial cones. It is possible to isolate and apply the solutions only within the central cone or between any two cones ignoring the rest of the field as there is no interaction between them. In order to find such solutions, the vector potential $\mathbf{\Pi}$ will be a function of $r$, $t$ and $\theta$. Although it is possible to find general solutions for the wave equation in spherical coordinates, we can by demanding self-similarity, design a set of multipole solutions that coincide with the Prendergast system. 

By setting $\mathbf{\Pi}=\Pi(v,\theta)\mathbf{\hat{z}}$ the wave equation is modified to the form:

\begin{eqnarray}
\Big[(v^{2}-1)(v^{2}\frac{\partial^{2}}{\partial v^{2}}+2v\frac{\partial}{\partial v})-\frac{1}{\sin \theta}\frac{\partial}{\partial \theta}\Big(\sin \theta \frac{\partial}{\partial \theta}\Big)\Big]\mathbf{\Pi}=0,
\end{eqnarray} 

and by the change of variable $u=\frac{1}{v}$ it simplifies to

\begin{eqnarray}
\Big[(u^{2}-1)\frac{\partial^{2}}{\partial u^{2}}+\frac{1}{\sin \theta}\frac{\partial}{\partial \theta}\Big(\sin \theta \frac{\partial}{\partial \theta}\Big)\Big]\mathbf{\Pi}=0.
\end{eqnarray}

When $\mathbf{\Pi}$ is independent of $\theta$ the solution is straightforward and equivalent to the case solved in the previous section. For $u>1$ we get

\begin{eqnarray}
\mathbf{\Pi}=m(u-1)\mathbf{\hat{z}},
\end{eqnarray}

and the results for the fields are the same as the ones described in equation (A13) for a linearly increasing dipole.

In the case of a multipole that is for $l>0$ it is possible to find analytical solutions for equation (B2) by separation of variables for $u>1$

\begin{eqnarray}
\mathbf{\Pi}=H(u-1)P_{l}(\mu)(u^{2}-1)P'_{l}(u)\mathbf{\hat{z}}.
\end{eqnarray}

where $H$ is Heaviside's step function and $\mu$ is $\cos \theta$. For $u<1$, zero satisfies equation (B2). In order to find the equivalence of these solutions to the Prendergast fields in empty space we will check if the fields defined in (A1) and (A2) obey the conditions set in expressions (1) and (6) 

\begin{eqnarray}
-\nabla \times \mathbf{\dot{\Pi}}=\frac{\mathbf{\hat{r}}}{u}\times (-\mathbf{\ddot{\Pi}}+\nabla(\nabla \cdot \mathbf{\Pi})).
\end{eqnarray}

Indeed (B5) reduces to (B2) after the differentiations are done; therefore, a field that obeys the self-similar wave equation, it will obey (1) and (6) as well. Thus the Hertzian multipole is equivalent to the multipole fields that obey the empty space Prendergast equation. 

To check that this solution (B4) holds not merely for $u>1$ and $u<1$ but also through $u=1$ we divide equation (B4) by the regular function $-(u+1)$ and integrate over the small region $-\epsilon<u-1<\epsilon$. After integrating by parts the left hand side gives:

\begin{eqnarray}
\Big(\Big\{\frac{\partial}{\partial u}\Big[(u^{2}-1)P_{l}'(u)H\Big](u-1)\Big\}^{1+\epsilon}_{1-\epsilon}{\nonumber}\\
-\Big[(u^{2}-1)P_{l}'(u)H\Big]^{1+\epsilon}_{1-\epsilon} & & {\nonumber}\\
-l(l+1)\int^{1+\epsilon}_{1-\epsilon}(u-1)P_{l}'(u)Hdu\Big)P_{l}(\mu) & & 
\end{eqnarray}

where $H$ stands for $H(u-1)$ so it is $1$ when $u=1+\epsilon$ and zero when $u=1-\epsilon$ and its derivatives are zero at both places. The first term reduces to $l(l+1)P_{l}(1+\epsilon)\epsilon$ and the second to $-\epsilon(2+\epsilon)P_{l}'(1+\epsilon)$ and the third reduces to the same integral without the $H$ but with the lower limit replaced by $1$. It is clearly $O(\epsilon^{2})$. Thus the whole expression is $O(\epsilon)$ and there is no $\delta$ function discontinuity in the equation at $u=1$. It is interesting to remark that had we divided equation (B4) by $1-u^{2}$ before integrating we have found the result

\begin{eqnarray}
\Big(\Big\{\frac{\partial}{\partial u}\Big[(u^{2}-1)P_{l}'(u)\Big]\Big\}^{1+\epsilon}-l(l+1)[P_{l}(u)]^{1+\epsilon}_{1}\Big)P_{l}(\mu)
\end{eqnarray}

which equals $l(l+1)P_{l}(1)=l(l+1)$, so the equation when divided by $1-u^{2}$ does have a $\delta$ function  discontinuity between the two sides at $u=1$! It is only the extra factor $u-1$ that saves the situation above. The derivation of equation (B2) from Maxwell's equations does not involve any multiplication by $u^{2}-1$.
Having demonstrated that our solution 

\begin{eqnarray}
\Pi(u,\mu)=P_{l}(\mu)(u^{2}-1)P_{l}'(u)H(u-1)
\end{eqnarray}

is a solution everywhere we recall that our starting equations are independent of the zero point of time so an explosion that starts at $t=\tau$ will be described by setting $u=c(t-\tau)/r$ rather than $ct/r$. Also, since our equations are the linear Maxwell equations, we can superpose solutions to obtain,

\begin{eqnarray}
\bar{\Pi}(ct,r,\mu)=\int^{\tau_{0}/2}_{\tau_{0}/2}\Pi(\frac{c(t-\tau)}{r},\mu)\frac{d\tau}{\tau_{0}}.
\end{eqnarray}

If $\Pi$ satisfies Maxwell's equations $\bar{\Pi}$, will too. The physics of this is that the explosion takes a finite time of order $\tau_{0}$ and occurs in a finite volume of order $(c\tau_{0})^{3}$ so we need to average the fields generated by a point explosion over radial distances $c\tau_{0}$.

It is not hard to perform this averaging because we are interested in regions where $u>>\frac{c\tau_{0}}{r}$ with $\tau_{0}$ small. We expand the function $(u^{2}-1)P_{l}'(u)$ with $u=\frac{c(t-\tau_{0})}{r}$ about $u=u_{0}$ where $u_{0}=\frac{ct}{r}$.

\begin{eqnarray}
(u^{2}-1)P_{l}'(u)=(u_{0}^{2}-1)P_{l}'(u_{0})+(u-u_{0})l(l+1)P_{l}(u_{0}) {\nonumber} \\
+\frac{1}{2}(u-u_{0})^{2}l(l+1)P_{l}'(u_{0})+O(u-u_{0})^{3},
\end{eqnarray}

where we used the fact that $P_{l}$ satisfies Legendre's equation. Now when the step in the Heaviside function in equation (B8) is at later times than those involved in the integral in (B9) we replace H by unity. So provided $r \leq c(t-\tau_{0}/2)$ we perform the integral in (B9) and find

\begin{eqnarray}
\bar{\Pi}=\Big[(u_{0}^{2}-1)P_{l}'(u_{0})+\frac{c^{2}\tau_{0}^{2}}{24r^{2}}l(l+1)P_{l}(u_{0})\Big]P_{l}(\mu);\\ 
u_{0}-1\geq\frac{c\tau_{0}}{2r} {\nonumber}.
\end{eqnarray}

However when $|t-r/c|<\tau_{0}/2$ the Heaviside function cuts into the range of integration in (B9). If we multiply equation (B10) by $H(u-1)$ we then find that the range of integration has its upper limit changed from $\tau_{0}/2$ to $t-r/c$. To evaluate $\bar{\Pi}$ we need the three integrals 

\begin{eqnarray}
I_{n}=\int_{-\tau_{0}/2}^{t-r/c}(u-u_{0})^{2}\frac{d\tau}{\tau_{0}}
\end{eqnarray}

for n=0, 1 and 2. These are readily evaluated:
\begin{eqnarray}
I_{0}=\Big(\frac{ct-r}{c\tau_{0}}+\frac{1}{2}\Big) \\
I_{1}=\frac{c\tau_{0}}{2r}\Big[\frac{1}{4}-\Big(\frac{ct-r}{c\tau_{0}}\Big)^{2}\Big]\\
I_{2}=\frac{c^{2}\tau_{0}^{2}}{3r^{2}}\Big[\frac{1}{8}+\Big(\frac{ct-r}{c\tau_{0}})^{3}\Big]
\end{eqnarray}

Notice that the variable $(ct-r)/(c\tau_{0})$ lies in the range $-1/2$ to $+1/2$ because $|t-r/c|\leq \tau_{0}/2$ in this region. Thus

\begin{eqnarray}
\bar{\Pi}=O\Big(\frac{c\tau_{0}}{r}\Big)^{3}+\Big[(u_{0}^{2}-1)P_{l}'(u_{0})I_{0}+l(l+1)P_{l}(u_{0})I_{1}+ & \nonumber\\
+\frac{1}{2}l(l+1)P_{l}'(u_{0})I_{2}\Big]P_{l}(\mu), |t-r/c|\leq\tau_{0}/2. &
\end{eqnarray}

As a further check we took $l=1$ and showed that this expression for $\bar{\Pi}$ does indeed obey equation (A13) which itself follows directly from Maxwell's equations. Thus our procedure of finding singular self-similar solutions and then averaging out the singularities is indeed giving non-singular solutions of Maxwell's equations even in the region over which the singularity is smoothed out. Of course in the new region (III) $r\geq c(t+\tau_{0}/2)$ there is no signal and $\bar{\Pi}$ is zero. Remembering that $u_{0}=ct/r$ we see that $\bar{\Pi}$ is given by 

\begin{eqnarray}
\bar{\Pi}=\Bigg\{ \begin{array}{ll} 
\Big[(u_{0}^{2}-1)+\frac{c^{2}\tau_{0}^{2}}{24r^{2}}l(l+1)\Big]P_{l}'(u_{0})P_{l}(\mu) & u_{0}-1> \frac{c\tau_{0}}{2r}\\
\{[(u_{0}^{2}-1)I_{0}+\frac{1}{2}l(l+1)I_{2}]P_{l}'(u_{0})+ & \\
+l(l+1)I_{1}P_{l}(u_{0})\}P_{l}(\mu) & |u_{0}-1|\leq \frac{c\tau_{0}}{2r} \\
0 & u_{0}-1<-\frac{c\tau_{0}}{2r}.
\end{array} 
\end{eqnarray}

It is easy to see that $\bar{\Pi}$ is continuous as $\big(\frac{ct-r}{c\tau_{0}}\big)=\pm 1/2$ at the edges of region (II), so $I_{0}$ and $I_{2}$ go from 1 and $\frac{c^{2}\tau_{0}^{2}}{12r^{2}}$ at the inner edge of region (II), to zero at the outer edge and $I_{1}$ is zero at both edges. Such a continuity is actually inevitable since even the unsmoothed $\Pi(u,\mu)$ is continuous since $(u^{2}-1)H(u-1)$ is continuous but furthermore the first derivative of $\Pi$ includes a Heaviside function discontinuity. The result of smoothing in $t$ ensures that the $\dot{\bar{\Pi}}$ is continuous but $\ddot{\bar{\Pi}}$ may have a Heaviside function corresponding to finite sudden changes in field strength at the edges of region (II). Even such jumps would be avoided had we used the gentler function $\frac{1}{30\tau_{0}^{5}}(\tau+\frac{\tau_{0}}{2})^{2}(\tau-\frac{\tau_{0}}{2})^{2}d\tau$ with $\tau$ in the range $-\tau_{0}/2$ to $+\tau_{0}/2$ in place of the simpler function $\frac{d\tau}{\tau_{0}}$ used above. The more complicated function has zero first derivatives at both ends of the range and would lead to continuous and differentiable field strengths everywhere but we felt that the much more complicated expression for the $I_{n}$ added unnecessary complexity. 

We notice that when $\tau_{0}$ is small, the smoothing produced a negligible modification of order $(c\tau_{0}/r)^{2}$ in region (I). In region (II) the changes are significant in that the $\delta$ function fields are removed and spread out over a region $\Delta r=c\tau_{0}$ around $r=ct$. The $I_{1}$ term is especially significant here, despite its falling to zero at the inner and outer edges of this region.

\bsp

\label{lastpage}


\begin{thebibliography}{}

\bibitem[\protect\citeauthoryear{{Aly}}{{Aly}}{1994}] {1994A&A...288.1012A}{Aly} J.J. 1994, \aap, 288, 1012A

\bibitem[\protect\citeauthoryear{{Bell} \& {Lucek}}{{Bell} \& {Lucek}}{1995}]
{1995MNRAS.277.1327B} {Bell} A.R., {Lucek} S.G. 1995 \mnras, 277, 1327B

\bibitem[\protect\citeauthoryear{{Coffey}, {Bacciotti}, {Ray}, {Eisl{\"o}ffel}, \& {Woitas}}{{Coffey}, {Bacciotti}, {Ray}, {Eisl{\"o}ffel}, \& {Woitas}}{2007}] {2007ApJ...663..350C}
{Coffey} D., {Bacciotti} F., {Ray} T. P., {Eisl{\"o}ffel} J., {Woitas} J. 2007 \apj, 663, C

\bibitem[\protect\citeauthoryear{{Contopoulos}}{{Contopoulos}}{1995}]
{1995ApJ...446...67C} {Contopoulos} J. 1995 \apj, 446, 67C

\bibitem[\protect\citeauthoryear{{De Villiers}, {Hawley}, {Krolik} \& {Hirose}}{{De Villiers}, {Hawley}, {Krolik} \& {Hirose}}{2005}] {2005ApJ...620..878D} {De Villiers} J.-P., {Hawley} J.F., {Krolik} J.H.\& {Hirose} S. 2005 \apj, 620, 878D

\bibitem[\protect\citeauthoryear{{Duncan} \& {Thompson}}{{Duncan} \& {Thompson}}{1992}]{1992ApJ...392L...9D} {Duncan} R.C., {Thompson} C., 1992, \apj, 392L, 9D

\bibitem[\protect\citeauthoryear{{Fendt} \& {Ouyed}}{{Fendt} \& {Ouyed}}{2004}]
{2004ApJ...608..378F} {Fendt} C., {Ouyed} R. 2004 \apj, 608, 378F

\bibitem[\protect\citeauthoryear{{Fendt}}{{Fendt}}{1997}] {1997A&A...319.1025F} {Fendt}, C. 1997 \aap, 319, 1025F

\bibitem[\protect\citeauthoryear{{Gammie}, {McKinney} \& {T\'oth}}{{Gammie}, {McKinney} \& {T\'oth}}{2003}] {2003ApJ...589..444G} {Gammie} C.F., {McKinney} J.C. \& {T\'oth} G., 2003 \apj, 589, 444G

\bibitem[\protect\citeauthoryear{{Grad} \& {Rubin}}{{Grad} \& {Rubin}}{1958}] {1958} {Grad} H., {Rubin} H. 1958, MHD Equilibrium in an Axisymmetric Toroid. Proceedings of the 2nd UN Conf. on the Peaceful Uses of Atomic Energy, Vol. 31, Geneva: IAEA p.190.

\bibitem[\protect\citeauthoryear{{Gourgouliatos}}{{Gourgouliatos}}{2008}] {2008MNRAS...385..875G} {Gourgouliatos} K.N., 2008 \mnras, 385, 875G

\bibitem[\protect\citeauthoryear{{Heger}, {Fryer}, {Woosley}, {Langer} \& {Hartmann}}{{Heger}, {Fryer}, {Woosley}, {Langer} \& {Hartmann}}{2003}] {2003ApJ...591..288H} {Heger} A., {Fryer} C.L., {Woosley} S.E., {Langer} N. \& {Hartmann} D.H. 2003 \apj, 591, 288H

\bibitem[\protect\citeauthoryear{{Jackson}}{{Jackson}}{1975}] {1975clel.book.....J} {Jackson} J.D. 1975, Classical Electrodynamics, Wiley21

\bibitem[\protect\citeauthoryear{{Komissarov}}{{Komissarov}}{2002}]{2002MNRAS.336..759K} {Komissarov} S.S. 2002 \mnras, 336, 759K


\bibitem[\protect\citeauthoryear{{Kouveliotou et al.}}{{Kouveliotou et al.}}{1998}] {1998Natur...3935..235K}
{Kouveliotou} C., 1998,	 \nat, 393, 235K

\bibitem[\protect\citeauthoryear{{Landau} \& {Lifshitz}}{{Landau} \& {Lifshitz}}{1975}] {1975ctf..book.....L} {Landau} L.D., {Lifshitz} E.M. 1975, The Classical Theory of Fields, Elsevier

\bibitem[\protect\citeauthoryear{{Le Blanc} \& {Wilson}}{{Le Blanc} \& {Wilson}}{1970}]
{1970ApJ...161..541L} {Le Blanc} J.M., {Wilson} J.R., 1970 \apj, 161, 541L

\bibitem[\protect\citeauthoryear{{Li}, {Chiueh} \& {Begelman}}{{Li}, {Chiueh} \& {Begelman}}{1992}] {1992ApJ...394..459L} {Li} Z.-Y., {Chiueh} T. \& {Begelman} M. 1992 \apj, 394, 459L

\bibitem[\protect\citeauthoryear{{Lovelace}}{{Lovelace}}{1976}] {1976Natur.262..649L} {Lovelace} R.V.E. 1976, \nat, 262, 649L

\bibitem[\protect\citeauthoryear{{Lynden-Bell}}{{Lynden-Bell}}{1978}] {1978PhyS...17..185L} {Lynden-Bell} D. 1978, \PhyS, 17, 185L

\bibitem[\protect\citeauthoryear{{Lynden-Bell} \& {Boily}}{{Lynden-Bell} \& {Boily}}{1994}] {1994MNRAS...267..146L}
{Lynden-Bell} D., {Boily} C. 1994, \mnras, 267, 146L

\bibitem[\protect\citeauthoryear{{Lynden-Bell}}{{Lynden-Bell}}{2006}] {2006MNRAS.369.1167L} {Lynden-Bell} D. 2006, \mnras, 369, 1167L 

\bibitem[\protect\citeauthoryear{{Mestel}}{{Mestel}}{1999}] {1999stma.book.....M} {Mestel} L. 1999, Cosmic Magnetism, Oxford 

\bibitem[\protect\citeauthoryear{{Murakami}, {Inoue}, {Nishimura} {et al.}}{{Murakami},{Inoue}, {Nishimura} {et al.}}{1991}] {1991Natur.350..592M} {Murakami} T., {Inoue} H., {Nishimura} J. {et al.},
1991 \nat, 350, 592M

\bibitem[\protect\citeauthoryear{{Ouyed}, {Pudritz} \& {Stone}}{{Ouyed}, {Pudritz} \& {Stone}}{1997}] {1997Natur.385..409O} {Ouyed} R., {Pudritz} R.E. \& {Stone} J.M. 1997 \nat, 385, 409O  

\bibitem[\protect\citeauthoryear{{Piran}}{{Piran}}{2005}] {2005RvMP...76.1143P} {Piran} T. 2005, \RvMP, 76, 1143P

\bibitem[\protect\citeauthoryear{{Prendergast}}{{Prendergast}}{2005}] {2005MNRAS.359..725P}{Prendergast} K.H. 2005, \mnras, 359, 725P

\bibitem[\protect\citeauthoryear{{Priest}}{{Priest}}{1984}] {1984smh..book.....P}{Priest} E.R. 1984, Solar Magnetohydrodynamics, Reidel, Dordrecht

\bibitem[\protect\citeauthoryear{{Rees} \& {M\'esz\'aros}}{{Rees} \& {M\'esz\'aros}}{2005}]
{2005ApJ...628..847R} {Rees} M.J. \& {M\'esz\'aros} P 2005 \apj, 628, 847R

\bibitem[\protect\citeauthoryear{{Sauty} \& {Tsinganos}}{{Sauty} \& {Tsinganos}}{1994}]
{1994A&A...287..893S} {Sauty} C. \& {Tsinganos} K. 1994 \aap, 287, 893S

\bibitem[\protect\citeauthoryear{{Shafranov}}{{Shafranov}}{1966}] {1966} {Shafranov} V.D. 1966, Plasma equilibrium in a magnetic field, Reviews of Plasma Physics, Vol. 2, New York: Consultants Bureau, p. 103

\bibitem[\protect\citeauthoryear{{Sherwin} \& {Lynden-Bell}}{{Sherwin} \& {Lynden-Bell}}{2007}] {2007MNRAS.378..409S}
{Sherwin} B.D., {Lynden-Bell} D. 2007, \mnras, 378, 409S

\bibitem[\protect\citeauthoryear{{Sommerfeld}}{{Sommerfeld}}{1952}]{1952} {Sommerfeld} A., 1952, Electrodynamics, New York 

\bibitem[\protect\citeauthoryear{{Spruit}}{{Spruit}}{2008}] {2008AIPC..983..391S} {Spruit} H. 2008, \Aipc, 983, 391S 

\bibitem[\protect\citeauthoryear{{Sturrock}}{{Sturrock}}{1994}] {1994ppai.book.....S} {Sturrock} P.A. 1994, Plasma Physics, Cambridge University Press

\bibitem[\protect\citeauthoryear{{Tchekhovskoy}, {McKinney} \& {Narayan}}{{Tchekhovskoy}, {McKinney} \& {Narayan}}{2008}] {2008MNRAS.388..551T} {Tchekhovskoy} A., {McKinney} J.C. \& {Narayan} R. 2008 \mnras, 388, 551T

\bibitem[\protect\citeauthoryear{{Uzdensky} \& {MacFadyen}}{{Uzdensky} \& {MacFadyen}}{2006}]
{2006ApJ...647.1192U} {Uzdensky} D.A. \& {MacFadyen} A.I, 2006 \apj, 647, 1192U

\bibitem[\protect\citeauthoryear{{Wolfson} \& {Low}}{{Wolfson} \& {Low}}{1992}]{1992ApJ...391..353W} {Wolfson} R., {Low} B.C. 1992 \apj, 391, 353W

\end{thebibliography}
\end{document}